\documentclass[3p]{elsarticle}

\usepackage[dvipsnames]{xcolor}
\usepackage[numbers]{natbib}
\usepackage{amssymb}
\usepackage[numbers]{natbib}
\usepackage{subcaption}
\usepackage{graphicx}
\usepackage{flafter}
\usepackage{gensymb} 
\usepackage{longtable}
\usepackage{float}
\usepackage{amsmath}
\usepackage{lineno}
\usepackage{setspace}

\journal{Nucl. Inst. and Meth. in Phys. Res. Sec. A}

\newcommand{\norm}[1]{\left\lVert#1\right\rVert}
\newcommand{\tripoli}{{\sc Tripoli-4}\textsuperscript{\textregistered}}
\newcommand{\Sab}[0]{S($\alpha$,$\beta$) }

\begin{document}

\begin{frontmatter}



\title{Improvement of Geant4 Neutron-HP package: from methodology to evaluated nuclear data library}


\author[label1]{L. Thulliez}\corref{cor1}
\ead{loic.thulliez@cea.fr}
\author[label2]{C. Jouanne}
\author[label1]{E. Dumonteil}

\cortext[cor1]{Corresponding author}

\affiliation[label1]{
            organization={IRFU, CEA, Universit\'e Paris-Saclay, 91191 Gif-sur-Yvette, France} 
            }
\affiliation[label2]{organization={Université Paris-Saclay, CEA, Service d’études des réacteurs et de mathématiques appliquées, 91191 Gif-sur-Yvette, France}
            }

\begin{abstract}
An accurate description of interactions between thermal neutrons (below 4 eV) and materials is key to simulate the transport of neutrons in a wide range of applications such as criticality-safety, reactor physics, compact accelerator-driven neutron sources, radiological shielding or nuclear instrumentation, just to name a few. While the Monte Carlo transport code Geant4 was initially developed to simulate particle physics experiments, its use has spread to neutronics applications, requiring evaluated cross-sections for neutrons and gammas between 0 and 20 MeV (the so-called neutron High Precision -HP- package), as well as a proper offline or on-the-flight treatment of these cross-sections. In this paper we will point out limitations affecting the Geant4 (version 10.07.p01) thermal neutron treatment and associated nuclear data libraries, by using comparisons with the reference Monte Carlo neutron transport code \tripoli, version 11, and we will present the results of various modifications of the Geant4 Neutron-HP package, required to overcome these limitations. Also, in order to broaden the support of nuclear data libraries compatible with Geant4, a nuclear processing tool has been developed (the code is available on a GitLab repository) and validated allowing the use of the code together with ENDF/B-VIII.0 and JEFF-3.3 libraries for example. These changes will be taken into account in an upcoming Geant4 release.
\end{abstract}


\end{frontmatter}


\section{Introduction}
\label{sec:Intro}

The past few decades have seen a tremendous increase in the use of large-scale simulations and of Monte Carlo particle transport codes to tackle the always-growing accuracy required by the diverse needs of neutron transport related problems, such as those met in nuclear industry (e.g reactor physics or criticality-safety studies), in accelerator physics, and for medical or hybrid applications (e.g. the design of compact accelerator driven neutron sources). The accuracy of these large-scale particle transport simulations is however ultimately conditioned by the quality of the nuclear data on which the codes rely as well as on the precision of the numerical methods they use. In particular, the accurate description of the neutrons slowing-down and thermalization process is of paramount importance. Most of the actual Monte Carlo neutron transport codes distinguish three regimes of interactions between neutrons and target nuclei, depending on the incoming neutron energy. For energies higher than a few hundred keV (fast spectrum), the neutrons see the target nuclei as a fixed point -without motion, which legitimates the use of 0 K cross-sections. 
For energies below a few hundred keV but greater than a few eV (epithermal spectrum), neutrons see the medium as a free gas of nuclei at thermal equilibium, with a Maxwellian distribution of velocities. In this approach, called the 'free gas approximation' \cite{Lux:268101}, the nuclear reaction occurs with the neutron-target relative velocity. This is the origin of the nuclear resonance Doppler broadening and the well known associated Doppler Broadening Rejection Correction method \cite{BECKER2009470}. For energies below a few eV (thermal spectrum), the neutron energy and wavelength are respectively comparable to the chemical bond energies and the medium inter-atomic distances. Consequently the neutron does not only interact with the nuclei but also with the material as a whole. The scattering kernels taking into account all the molecular/material effects, known as S($\alpha$,$\beta$) or thermal scattering law (TSL), are available in evaluated nuclear data libraries such as ENDF/B-VII.1 \cite{Chadwick2011}, ENDF/B-VIII.0 \cite{Brown2018}, JEFF-3.3 \cite{JEFF-TSL2020}, etc. They are based on molecular dynamics calculations and experimental measurements. If molecular effects can be ignored or if TSL data are not available, the free gas approximation is used. The accurate description of the target nucleus thermal motion and exhaustive TSL data libraries are therefore keys to grasp slowing down and thermalization characteristics of thermal and cold neutrons in various materials.

To this aim, the nuclear industry and nuclear engineering laboratories have been developing and using for decades dedicated Monte Carlo neutron transport codes, relying on evaluated nuclear data and subsequent nuclear data processing and treatment. Among these reference codes for neutronics, MNCP (version 5~\cite{mcnp5} and 6~\cite{mcnp6}), SCALE~\cite{scale}, SERPENT~\cite{serpent}, MORET (version 5~\cite{moret5} and 6~\cite{moret6}) or \tripoli~ \cite{tripoli4} are validated and qualified using inter-code comparisons (see for instance \cite{intercomp}) and large qualification data bases (using either integral or differential experiments). Combined with a large user community, these codes hence present coherent results in their respective qualification domains. In a different context, the open-source Monte Carlo code Geant4 \cite{geant, Allison2016} was originally designed for high-energy physics and was extended a few years ago to transport low energy neutrons (below $20$ MeV) using the Neutron High Precision -HP- package. Improvements are continuously made to this package to increase its precision such as described in \cite{Mendoza2014,Hartling2018,Tran2018a}. However, large discrepancies still remain when Geant4 is compared to neutronics reference codes  and experiments. A quick review of the Neutron-HP package shows various shortcomings of the code, related either to the handling of resonances in general (in the unresolved and resolved regions) or to the handling of the free gas approximation itself \cite{Mendoza2018} and of TSL data \cite{Tran2018a} with potential consequences in the epithermal and thermal energy domains for all materials. While being out of the scope of this paper, it might be good to know that using Geant4 in a context of reactor physics (with heavy structure materials and fissile media) might require ad-hoc developments for a proper handling of resonances, specifically in the unresolved resonance region (use of the so-called probability tables instead of averaged cross sections which are used at present) and in the resolved resonance region (use of the Doppler Broadening Rejection Correction (DBRC) method \cite{Becker2009} which should lead to sensitive improvements of the elastic scattering kernel estimation close to neutron resonances).

This present works therefore focuses on the inconsistencies related to the free gas approximation and TSL data, and proposes developments within the Geant4 Neutron-HP package to correct them. It is organized as follows: in Section \ref{sec:Methods} the methodology is introduced, based on the use of \tripoli~as a reference code and on the definition and use of two simple neutron transport benchmarks (the homogeneous sphere and the thin cylinder). In Section \ref{sec:SVT}, the Geant4 free gas approximation implementation is revised with the "Sampling of the Velocity of the Target nucleus" (SVT) algorithm \cite{Coveyou1956} used in \tripoli. Its description and impact on Geant4 predictions are evaluated and compared to \tripoli. In Section \ref{sec:TSLGeant4} the corrections applied to the Geant4 TSL data treatment are presented along with their impact on computing time. A nuclear data processing tool suited to the production of updated and new TSL data is presented together with its validation using \tripoli. Therefore instead of being restricted to use TSL data from ENDF/B-VII.1 (dating back to 2011) in the latest Geant4 version, new libraries can be used such as ENDF/B-VIII.0 and JEFF-3.3.

\section{Methods: comparing Geant4 to \tripoli~on two benchmarks}
\label{sec:Methods}

\tripoli~is a continuous-energy radiation transport Monte Carlo code developed since the mid-1990s at CEA-Saclay and devoted to shielding, reactor physics with depletion, criticality-safety and nuclear instrumentation for both fission and fusion systems. It is used as a reference code by the main French nuclear companies~\cite{tripoli4} and benefits from a very large verification and validation database gathering more than 1000 experimental benchmarks (ICSBEP benchmarks~\cite{icsbep}, reactor physics experiments~\cite{blaise}) as well as many comparisons to the US reference Monte Carlo code MCNP (see for instance \cite{fausser}). It is qualified for various applications (criticality-safety, burn-up credit, reactor physics, etc.), closely following the recommendations of the French nuclear safety authority for nuclear safety demonstrations~\cite{asnguide28}.
In the following, Geant4 version 10.07.p01 (G4) will be compared to a recent release of \tripoli~version 11 (T4).

In order to validate and to compare the Geant4 accuracy against \tripoli, we will resort to two simple benchmarks designed to grasp inconsistencies in the treatment of scattering kernels at thermal and epithermal energies. These inconsistencies will be pointed out by calculating Geant4 relative errors with respect to \tripoli, highlighting differences through plots giving the (Geant4-Tripoli4)/Tripoli4 ratio versus the neutron incoming energy. 
\newline \indent
The 'microscopic' benchmark, that we will refer to as the "thin-cylinder benchmark", allows to finely probe the spectral and angular characteristics of the neutron having undergone exactly one collision \cite{Mendoza2014}. It consists of a very thin cylinder having a 1 $\mu$m radius and 2 m length. A mono-energetic neutron beam is sent along the cylinder axis. Since the cylinder radius is small compared to the neutron mean free-path, after one collision the neutron leaves the cylinder, granting access to the kinetics of the collision through ad hoc tallies of the energy and angle of the scattered neutron. In this benchmark the initial neutron energy is set to 10$^{-8}$ MeV.
\newline \indent
The 'macroscopic benchmark', referred to as the "sphere benchmark", allows investigating the accuracy of the neutron slowing-down and thermalization. It consists of a simple sphere made of a given material with a mono-energetic and isotropic neutron source placed at its center. Only the neutron flux inside the sphere is tallied. The initial neutron energy is set to 10 eV in order to probe the end of the slowing-down process, so as to investigate the transition between use of the nuclear cross-section above 4 eV and the TSL below 4 eV, and to probe the TSL.

\section{Free gas approximation - SVT method}
\label{sec:SVT}

\begin{figure}[htbp]
    \begin{center}
    \includegraphics[scale=0.5]{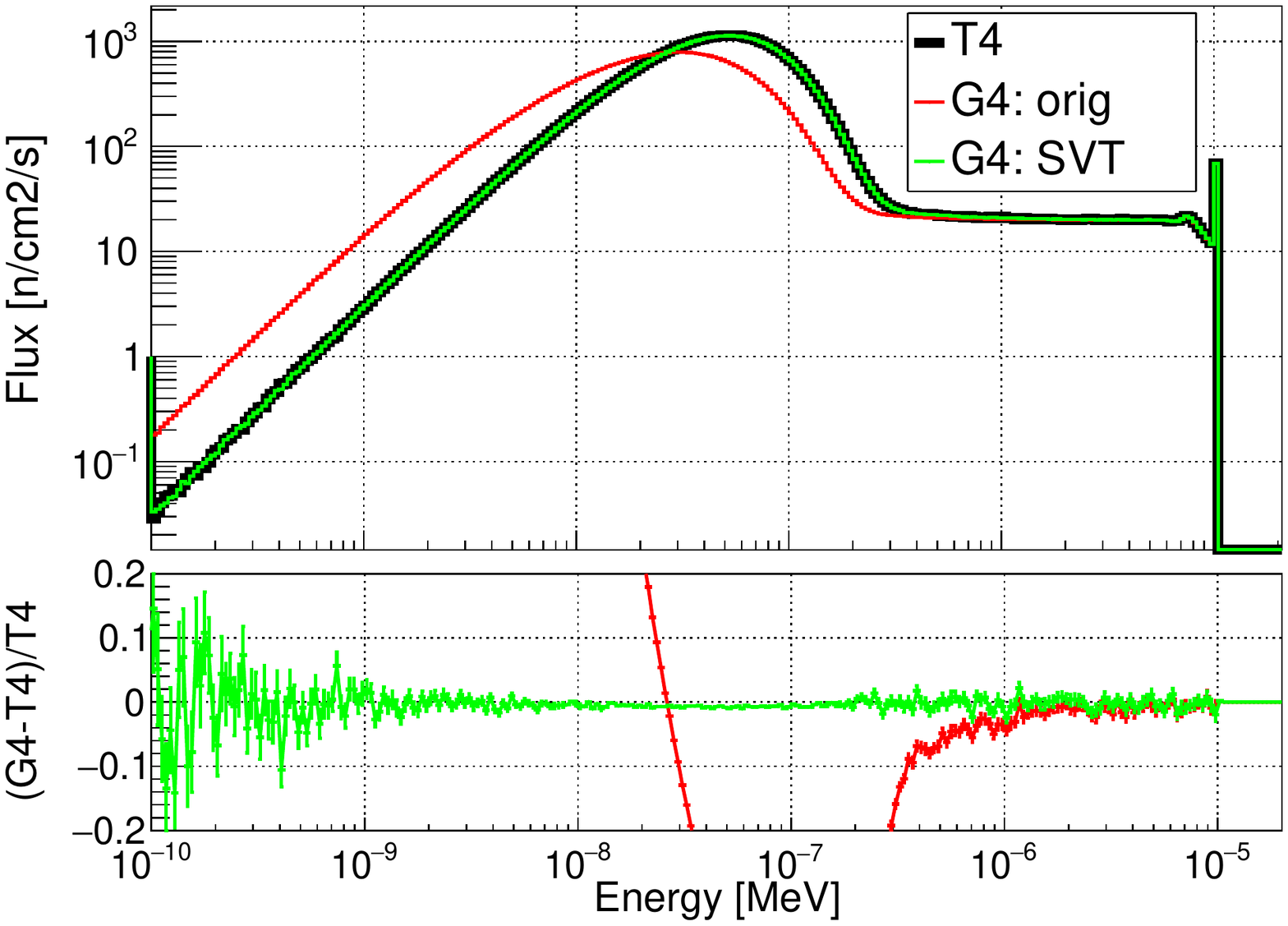}
	\end{center}
	  \caption{\label{fig:C12_sphereBenchmark} Neutron flux obtained with the sphere benchmark with the ENDF/B-VII.1 library for a $^{12}$C free gas medium (top plot) and their relative differences using \tripoli~as the reference (bottom plot), for Geant4 original algorithm (red curve), Geant4 modified algorithm (green curve) and \tripoli~(black curve).}
\end{figure}

As mentioned in numerous articles \cite{Mendoza2014,Mendoza2018}, one of the long standing issues related to the use of Geant4 Neutron-HP package is connected to the observation of large discrepancies between Geant4 and reference codes, originating from the Geant4 implementation of the free gas approximation and affecting tallies below 1 eV for the vast majority of isotopes. In Geant4, the three target velocity components are sampled from a Maxwellian distribution and then the velocity is accepted with a given probability. However, as can be observed in Figure~\ref{fig:C12_sphereBenchmark} comparing predictions for $^{\text{12}}$C material using \tripoli~(black) and the Geant4 original algorithm (red), discrepancies greater than 100 \% are visible. The conclusion is that Geant4 algorithm does not respect the thermal averaged reaction rate as in reference codes using the "Sampling of the Velocity of the Target nucleus" (SVT) algorithm to this end \cite{Coveyou1956}. When an epithermal neutron with a velocity $v_{n}$ is transported in a medium at a temperature $T$, it sees the material as a free gas of nuclei having a Maxwellian velocity distribution $\mathcal{M}(\vec{v}_T)$. The nuclear reaction uses an energy deduced from the neutron/target relative speed $\vec{v}_{R}=\vec{v}_{n}-\vec{v}_{T}$. To compute the kinematics of the elastic reaction, the target velocity $\norm{\vec{v_{T}}}$ and the collision angle (cos$\theta$=$\mu$) have to conserve the thermal averaged reaction rate given by: 

\begin{equation}\label{eq:ThermalAverageReactionRate}
 v_{n} \bar{\sigma}(v_{n}, T)= \int d\vec{v}_{T}\sigma(v_{r})\mathcal{M}(\vec{v}_T)
\end{equation}
\noindent 
where $\bar{\sigma}(v_{n}, T)$ is the averaged microscopic cross section. The ($\norm{\vec{v_{T}}}$, $\mu$) pair is sampled from the following joint probability distribution:

\begin{equation}\label{eq:pdfTarget}
    p(v_{T},\mu) = \frac{4\sigma_{s}}{\sqrt{\pi}C}\norm{\vec{v}_{n}-\vec{v}_{T}} \beta^{3}v_{T}^{2}e^{-\beta^{2}v_{T}^{2}}
\end{equation}
\noindent
where $\beta=\sqrt{\frac{M}{2k_{B}T}}$, $M$ is the target mass, $k_{B}$ the Boltzmann constant and $C$ a normalisation constant. This equation can be re-written as:

\begin{equation}\label{eq:jointDistrib_f}
    p(v_{T},\mu) \propto \underbrace{\frac{\sqrt{v_{n}^{2}+v_{T}^{2}-2v_{n}v_{T}\mu}}{v_{n}+v_{T}} }_{\text{(A)}}    \underbrace{(v_{n}+v_{T})\beta^{3}v_{T}^{2}e^{-\beta^{2}v_{T}^{2}}}_{\text{(B)}}
\end{equation}

\begin{figure}[htbp]
    \begin{center}
	\includegraphics[scale=0.5]{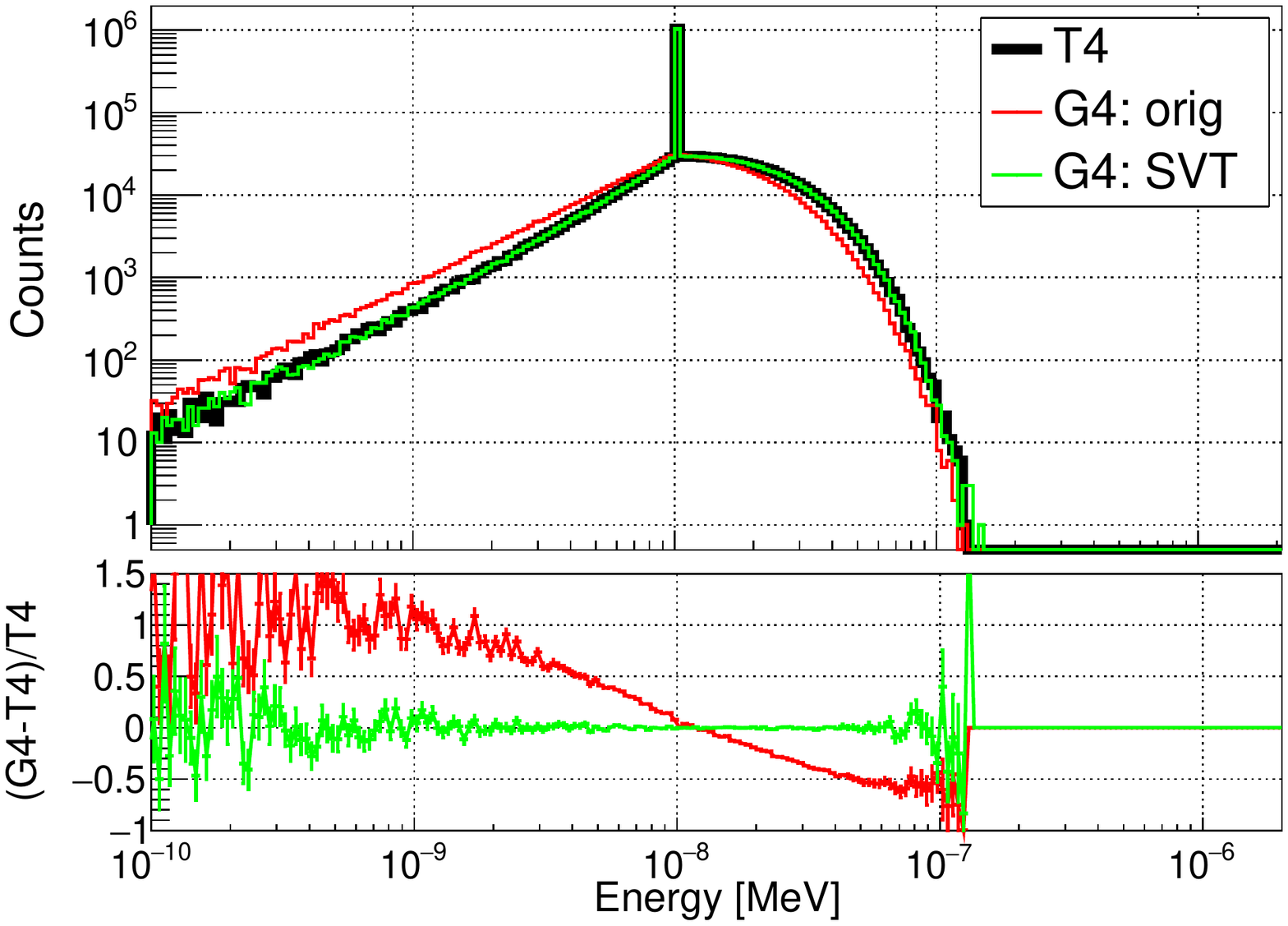}
	\end{center}
	\caption{\label{fig:C12_thinCylinderBenchmark_energy} Scattered neutron energy spectrum obtained with the thin cylinder benchmark with the ENDF/B-VII.1 library for a $^{12}$C free gas medium (top plot) and their relative differences using \tripoli~as the reference (bottom plot) for Geant4 original algorithm (red curve), Geant4 modified algorithm (green curve) and \tripoli~black curve).}
\end{figure}

\begin{figure}[htbp]
    \begin{center}
	\includegraphics[scale=0.5]{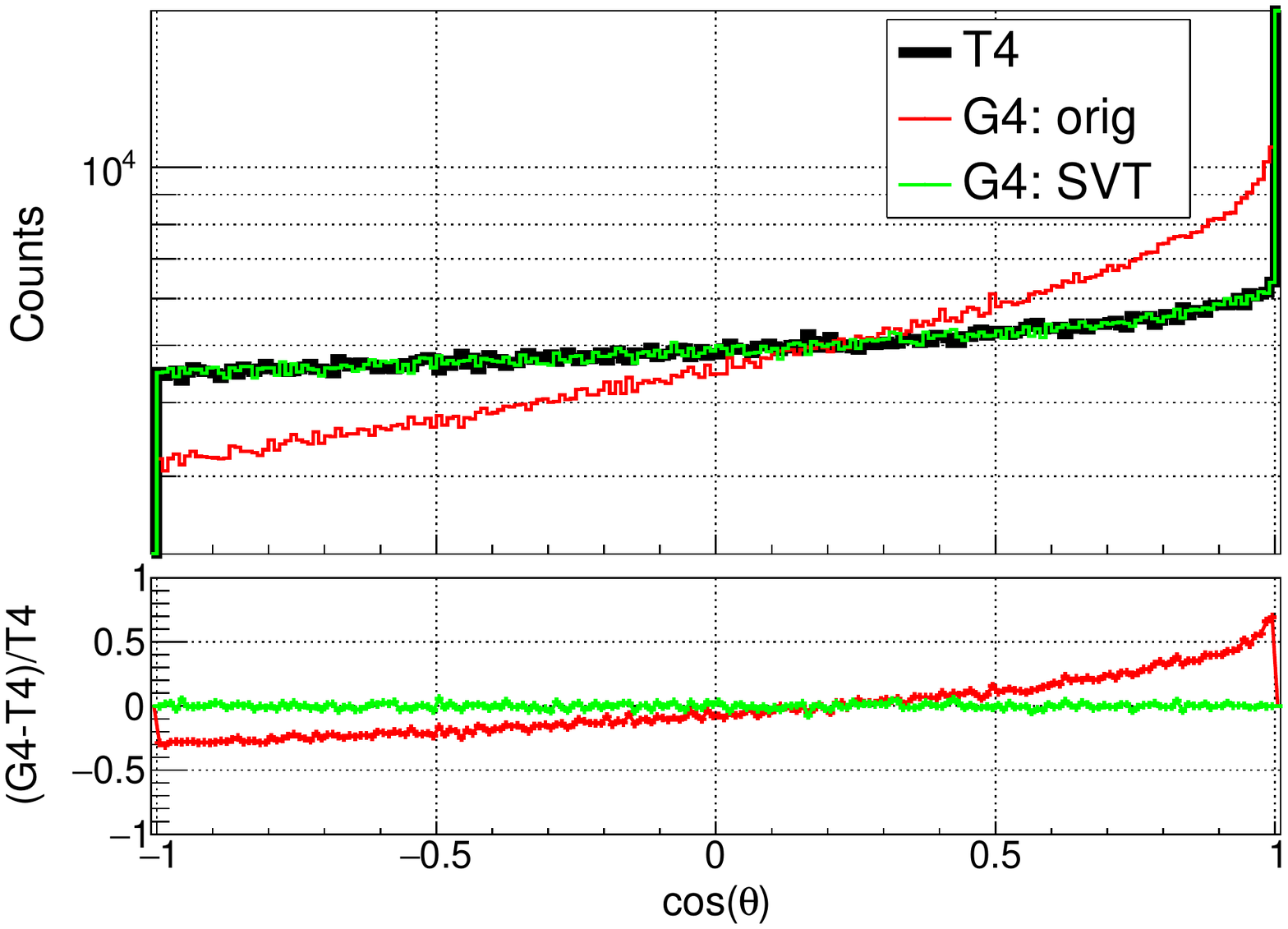}
	\end{center}
	\caption{\label{fig:C12_thinCylinderBenchmark_cosTheta} Scattered neutron cosinus angle obtained with the thin cylinder benchmark with the ENDF/B-VII.1 library for a $^{12}$C free gas medium (top plot) and their relative differences using \tripoli~as the reference (bottom plot) for Geant4 original algorithm (red curve), Geant4 modified algorithm (green curve) and \tripoli~(black curve).}
\end{figure}

The ($\norm{\vec{v_{T}}}$, $\mu$) pair is sampled from equation \ref{eq:jointDistrib_f} in three steps: (1) $\mu$ is sampled uniformly from [-1,1], (2) $\norm{\vec{v_{T}}}$ in equation \ref{eq:jointDistrib_f} term B is sampled using algorithms detailed in \cite{Everett1983} and (3) the pair ($\norm{\vec{v_{T}}}$, $\mu$) is accepted according to the probability defined by equation \ref{eq:jointDistrib_f} term A. If not, the routine is performed again. Once a pair is accepted, the kinematics of the elastic reaction is completely defined. 
This SVT algorithm has been implemented in the \textit{GetBiasedThermalNucleus} method of the \textit{G4Nucleus} class. 
Figure \ref{fig:C12_sphereBenchmark} presents its results for the $^{\text{12}}$C sphere benchmark (green curve). The discrepancies between \tripoli~and the Geant4 SVT algorithm decrease down to less than 1 \% (the original algorithm discrepancies were more than 100\%). The improvement brought by the SVT method can be also seen with the thin-cylinder benchmark presented in Figures \ref{fig:C12_thinCylinderBenchmark_energy} and \ref{fig:C12_thinCylinderBenchmark_cosTheta} where the differences between Geant4 and \tripoli~are lowered to less than 1 \% with a statistical uncertainty of $\pm$0.5 \%, in region of 8 meV to 200 meV. 
\newline \indent
While the implementation of the SVT algorithm within Geant4 seems therefore to solve long-standing issues related to the use of the Neutron-HP package, it is important to note that this method assumes that the cross-section is constant (this is necessary to go from equation \ref{eq:ThermalAverageReactionRate} to \ref{eq:pdfTarget}) over the energy range covered by the relative velocity computed for a given neutron velocity. Therefore this assumption holds true for nuclei with no resonance in the epithermal region, \textit{i.e.} for light and medium mass nuclei. For heavy nuclei such as uranium, the presence of resonances in the epithermal region induces large cross-section variations. Consequently the assumption used by the SVT algorithm breaks down and the so-called Doppler Broadening Rejection Correction (DBRC) algorithm needs to be used instead \cite{Becker2009, zoiadbrc1}. This has not been done in the present work since it requires deeper modifications of the Geant4 code. 

\section{Geant4 thermal scattering kernel treatment}
\label{sec:TSLGeant4}

In the thermal energy domain, the free gas approximation often breaks down because the neutron energy and wavelength are respectively comparable to chemical bond energies and to inter-atomic distances. The molecular interactions are taken into account \textit{via} evaluated thermal scattering kernels known as \Sab, where $\alpha$ and $\beta$ are respectively the dimensionless momentum and energy transfer \cite{Squires2012}. \Sab data are computed from the phonon density of states (e.g. vibration or rotation) which are produced by molecular dynamics codes (e.g. GROMACS or VASP \cite{gromacs,Kresse1996}) or experiments (e.g. see \cite{neatsl} for a compilation of experiments on the topic). The scattering kernel can be decomposed in a coherent and in an incoherent part. The coherent part depends on the position correlations at different times of the same or neighboring atoms having the same scattering length equivalent to the average scattering length of the atoms in the system. This gives rise to interference effects. Basically, to have a coherent part, the medium should have a structure, and hence should be a crystal. The incoherent part comes from the position correlation for the same atoms at different times with a scattering length different from the averaged scattering length. This can be encountered in every solid, liquid or gas.
Then the coherent or incoherent part can be split in an elastic and an inelastic part. In an inelastic interaction the neutron energy does change. The neutron can lose (down-scattering) or gain (up-scattering) energy respectively with phonon excitation and de-excitation in a solid, with vibrational, rotational or translational molecular excitation or deexcitation or by target recoil if it is not heavy. On the other hand the interaction is elastic if no energy is transferred to the recoil molecule or solid because of its infinite mass compared to the neutron mass. A coherent elastic process will lead to peaks in the cross-section if the Bragg rule is satisfied.
\newline 
TSL data are provided through evaluated nuclear data libraries in the ENDF-6 format, and are not directly usable by neutron transport codes. In fact, they are represented as tabulated scattering kernels \Sab. Therefore the kernels \Sab need to be processed to get double differential cross-sections $\frac{d^{2}\sigma}{d\Omega dE'}$ with the NJOY code \cite{Macfarlane2017} for example.
Often the coherent inelastic process is neglected during the construction of the \Sab, this is called the incoherent approximation (in certain conditions this could fail \cite{JEFF-TSL2020}). This approximation is used in the LEAPR module of NJOY. Therefore the neutron/medium interactions are split in coherent elastic, incoherent elastic and incoherent inelastic processes.

\subsection{New thermal scattering kernels: processing tool and validation}
\label{subsec:TSLprocessing}

From 2011 and up to now, only ENDF/B-VII.1 TSL data were available in Geant4, even if ENDF/B-VIII.0 or JEFF-3.3 have released updated and new TSL evaluations in the past years. In order to have access to the latest TSL evaluations and to study the impact of nuclear data processing parameters on Monte Carlo simulations and in particular in Geant4, a processing tool has been developed and validated against \tripoli. In this work NJOY-2016 \cite{Macfarlane2017} is used to process \Sab. The code is available on the following GitLab repository \cite{GitLab}.

\subsubsection{Processing tool}

A processing tool based on NJOY allowing to convert \Sab from ENDF file to Geant4 TSL files has been written. The first step is to generate the double differential cross-section and the second is to format the NJOY output file into Geant4 TSL files.
\newline
With NJOY the nuclear cross-section is Doppler broadened with the BROADR module at the processed temperature. This ensures the continuity between nuclear cross-sections (above 4 eV) and thermal scattering cross-sections (below 4 eV) at the energy cut-off equal to 4 eV in Geant4 (\tripoli's cut-off is 4.95 eV). Then the THERMR module is used to transform TSL data with user specifications related to the energy reconstruction tolerance ($tol$) and the number of equi-probable scattering cosine angle ($N_{\mu}$). The cross-section binning is built by NJOY to allow linear interpolation between two points within the tolerance $tol$. Hartling et al. \cite{Hartling2018} have shown that the parameters $tol$=0.001 and $N_{\mu}\geq$20 are required to accurately predict for example neutron transmission coefficients (see Fig. 7 and 8 in \cite{Hartling2018}). Following their recommendations, in this work $tol$=0.001 and $N_{\mu}$=32 are chosen. The THERMR ouput file is in a PENDF format which is transformed to be processed by Geant4. Schematically the total cross-section given by the MF=3 file are placed in the \textit{CrossSection} directories (Coherent, Incoherent, Inelastic), while final states from MF=6 file are placed in the \textit{FS} directories (Incoherent, Inelastic). The exception is made for the coherent elastic final state which are directly taken from the MF=7 (MT=2) evaluated data file which is processed by NJOY, because it already represents final states. 
\newline
\tripoli~has been chosen as a reference code partly because it also uses the THERMR output file to deal with TSL data. Therefore the same NJOY parameters are used in Geant4 and \tripoli~TSL data processing. 

\subsubsection{Validation}
\begin{figure}[H]
    
    \centering
     \begin{subfigure}[b]{0.49\textwidth}
         \centering
         \includegraphics[width=\textwidth]{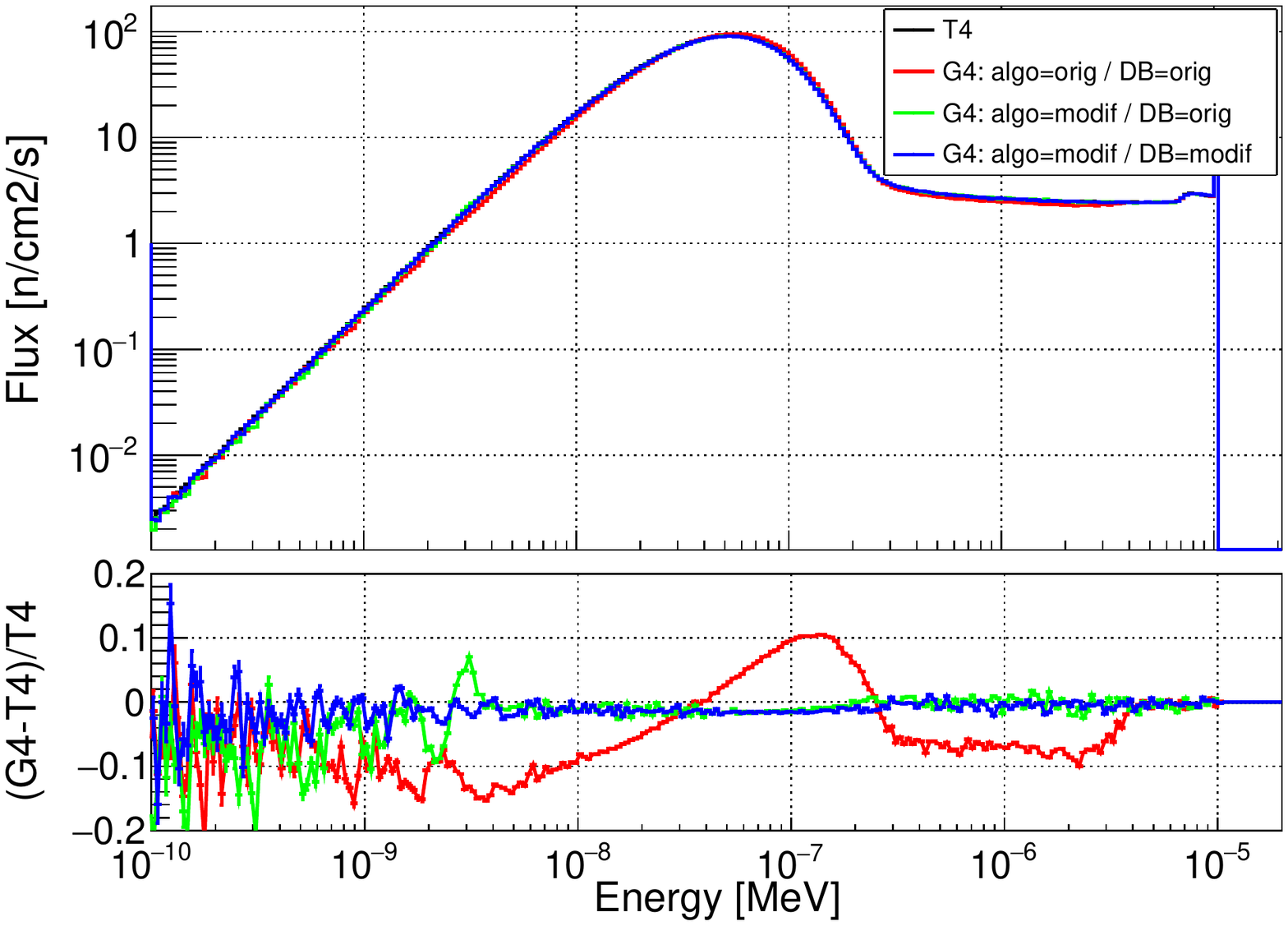}
         \caption{ENDF/B-VII.1 - CH$_2$ with HinCH$_2$ TSL - 296K}
     \end{subfigure}
     \hfill
     \begin{subfigure}[b]{0.49\textwidth}
         \centering
         \includegraphics[width=\textwidth]{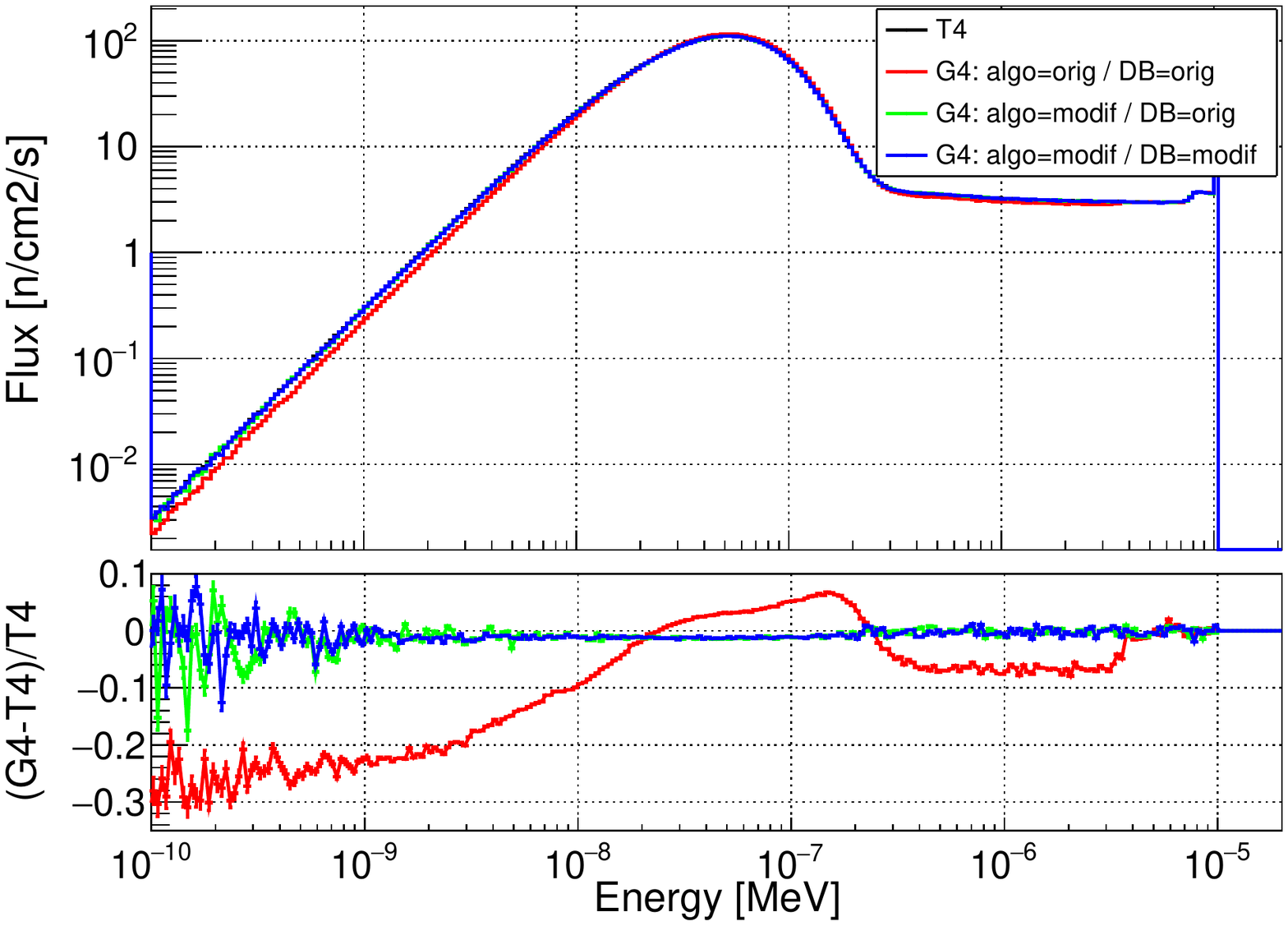}
         \caption{ENDF/B-VII.1 - H$_2$O with HinH$_2$O TSL - 294K}
     \end{subfigure}
     \hfill
     \begin{subfigure}[b]{0.49\textwidth}
         \centering
         \includegraphics[width=\textwidth]{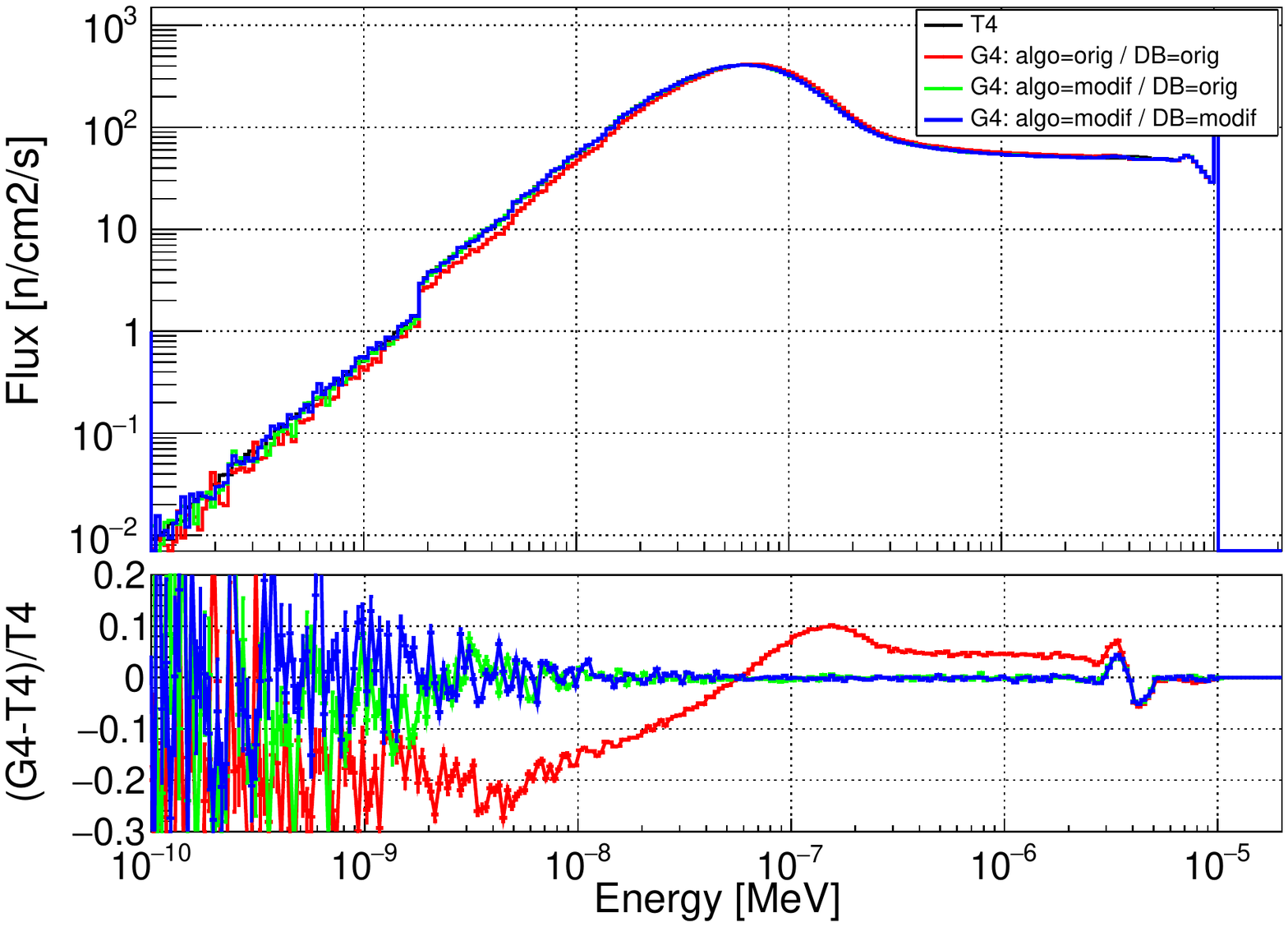}
         \caption{ENDF/B-VII.1 - Graphite TSL - 296K}
     \end{subfigure}
     \hfill
     \begin{subfigure}[b]{0.49\textwidth}
         \centering
         \includegraphics[width=\textwidth]{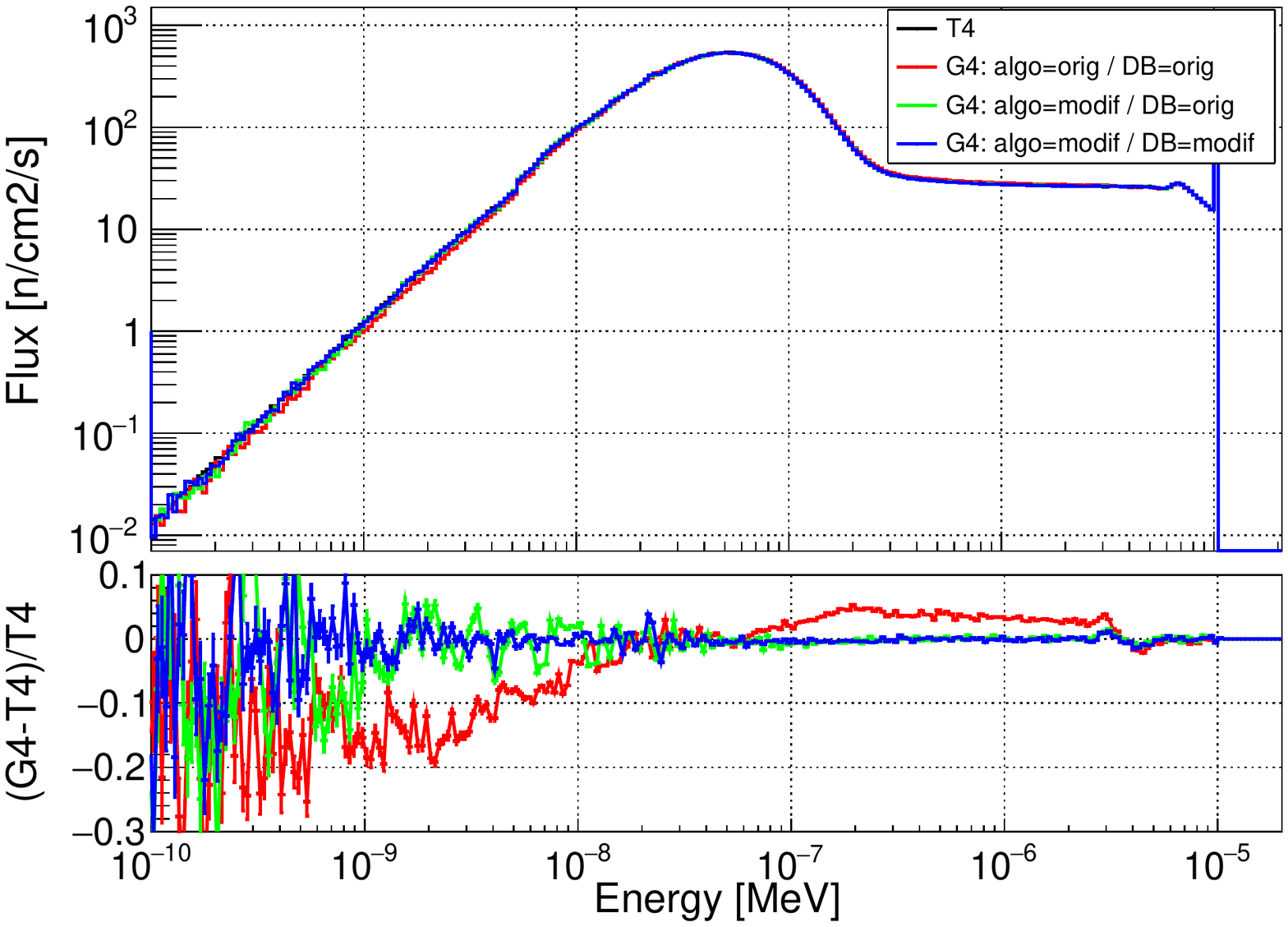}
         \caption{ENDF/B-VII.1 - Be metal TSL - 294K} 
     \end{subfigure}
     \hfill
     \begin{subfigure}[b]{0.49\textwidth}
         \centering
         \includegraphics[width=\textwidth]{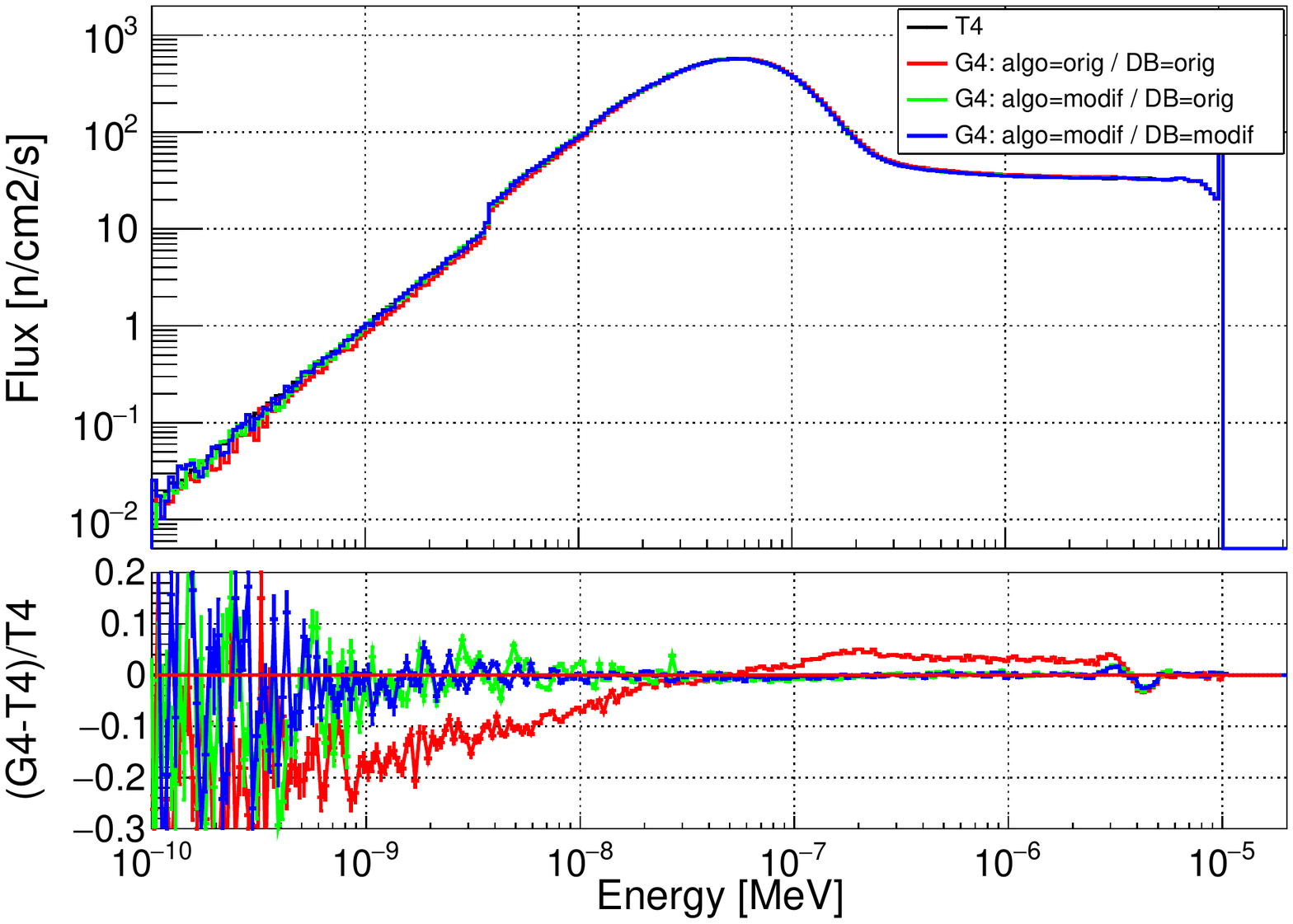}
         \caption{ENDF/B-VII.1 - BeO with BeinBeO and OinBeO TSL - 294K} 
     \end{subfigure}    
     \hfill
     \begin{subfigure}[b]{0.49\textwidth}
         \centering
         \includegraphics[width=\textwidth]{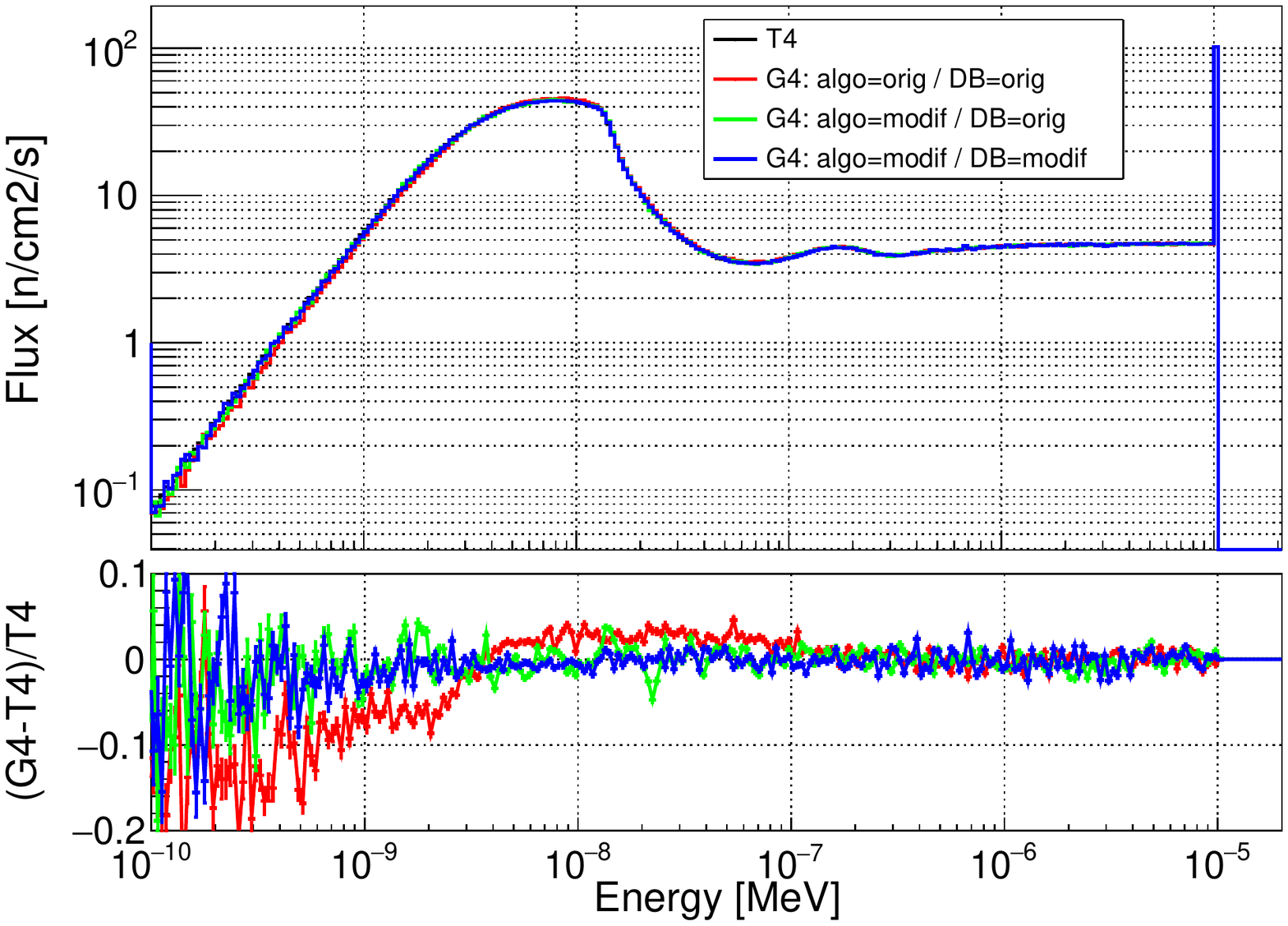}
         \caption{ENDF/B-VII.1 - para-H$_2$ TSL - 20K}
     \end{subfigure}
     
     \caption{\label{fig:B7_sphere_all} Neutron flux obtained with the sphere benchmark with the ENDF/B-VII.1 library for different medium described by TSL data (top plot) and their relative differences using \tripoli~as the reference (bottom plot), for Geant4 original algorithm (red curve), Geant4 modified algorithm (green curve), Geant4 modified algorithm and reprocessed TSL data (blue curve)  and \tripoli~(black curve).}
        
\end{figure}

\begin{figure}[H]

    \centering
     \begin{subfigure}[b]{0.49\linewidth}
         \centering
         \includegraphics[width=\linewidth]{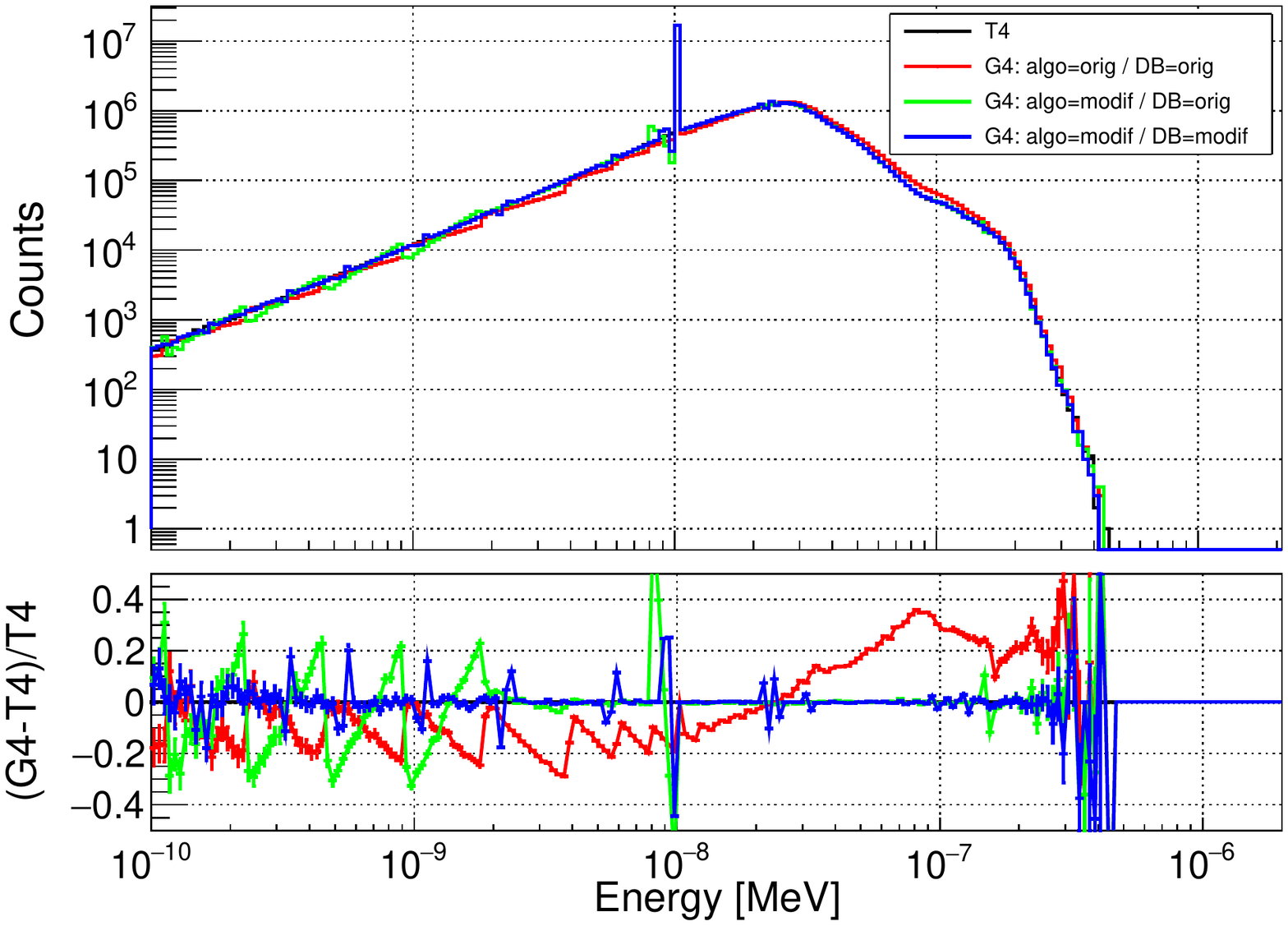}
         \caption{ENDF/B-VII.1 - CH$_2$ with HinCH$_2$ TSL - 296K}
     \end{subfigure}
     \hfill
     \begin{subfigure}[b]{0.49\linewidth}
         \centering
         \includegraphics[width=\linewidth]{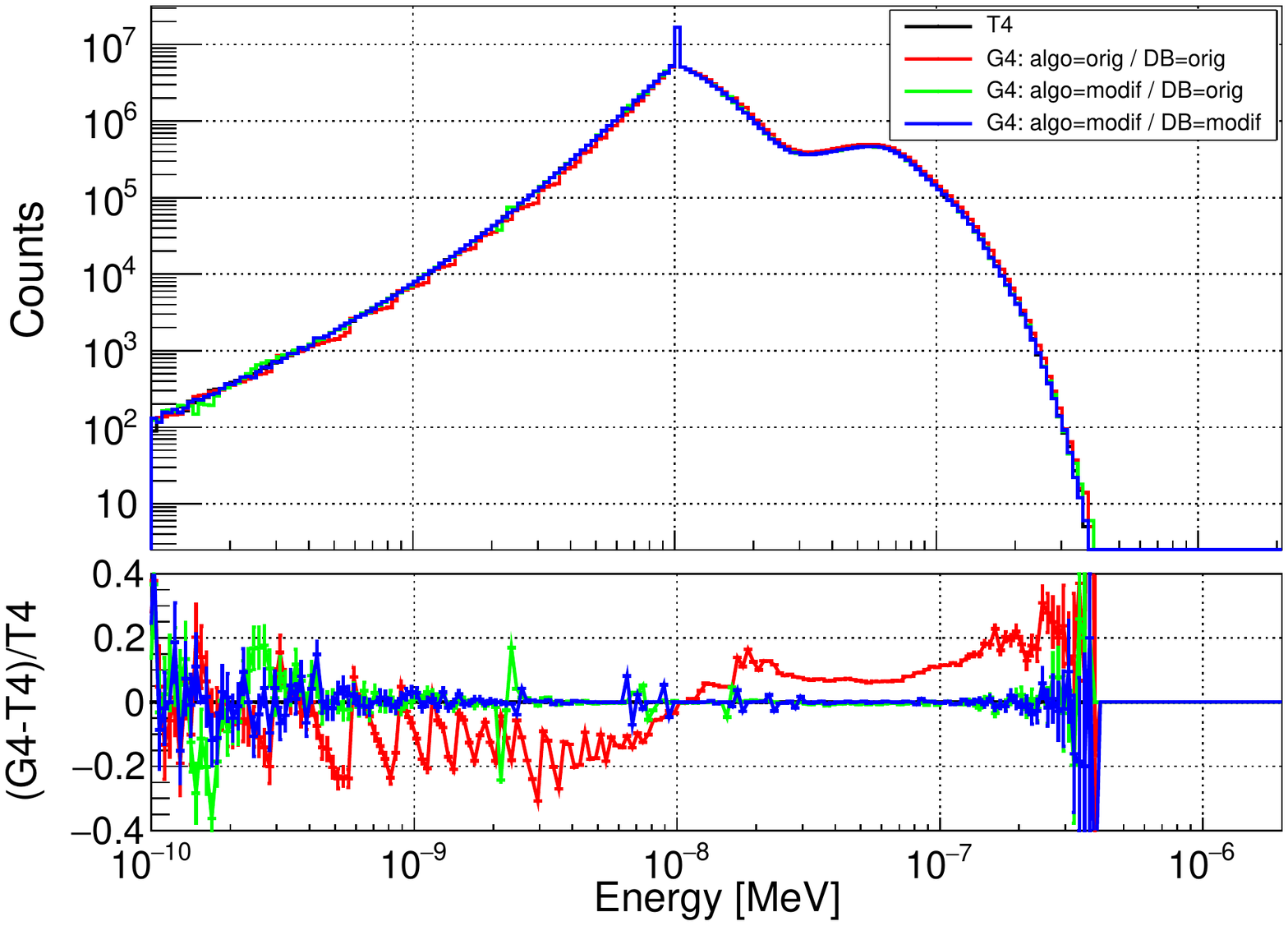}
         \caption{ENDF/B-VII.1 - H$_2$O with HinH$_2$O TSL - 294K}
     \end{subfigure}
     \hfill
     \begin{subfigure}[b]{0.49\linewidth}
         \centering
         \includegraphics[width=\linewidth]{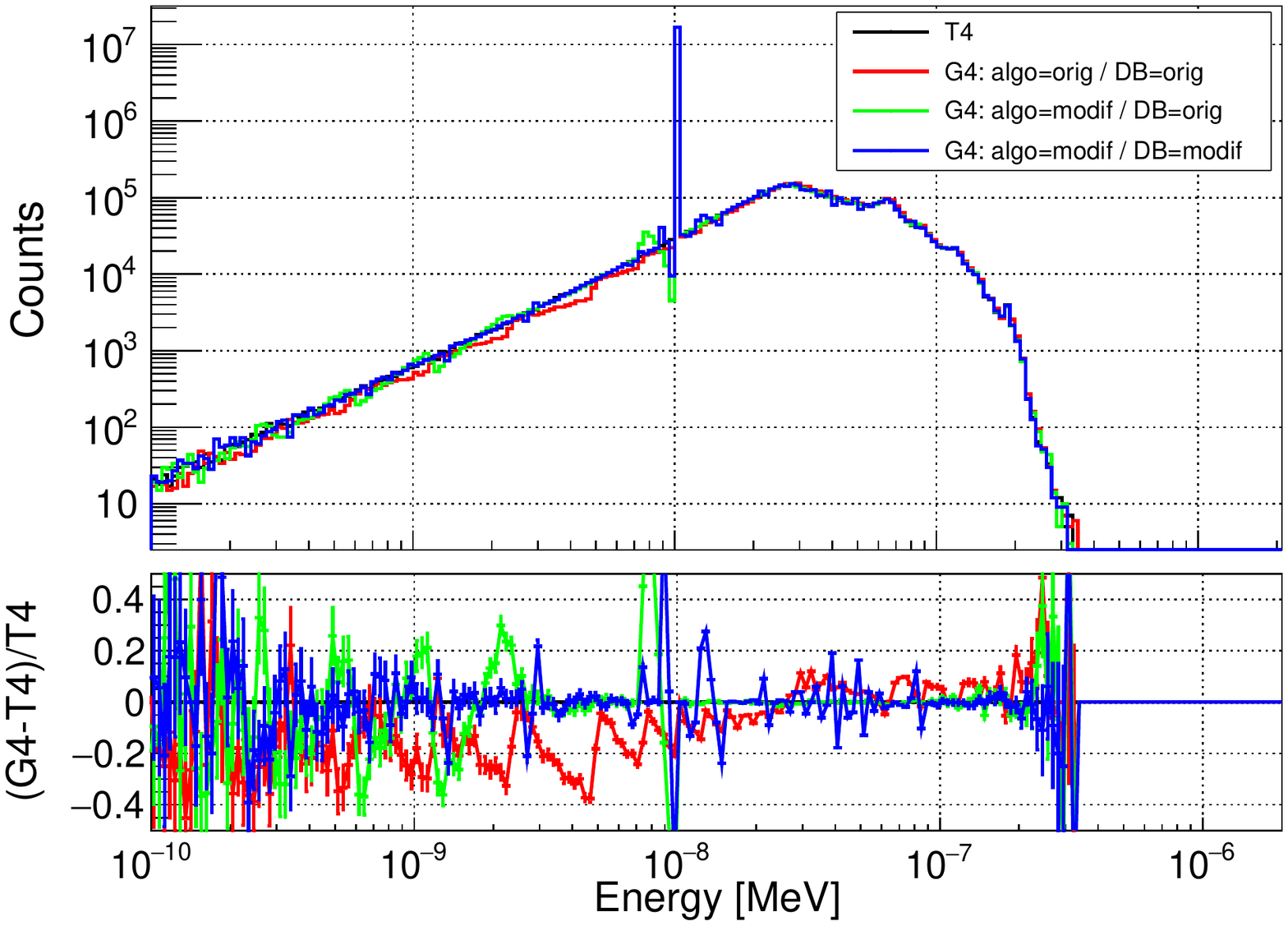}
         \caption{ENDF/B-VII.1 - Graphite TSL - 296K}
     \end{subfigure}
     \hfill
     \begin{subfigure}[b]{0.49\linewidth}
         \centering
         \includegraphics[width=\linewidth]{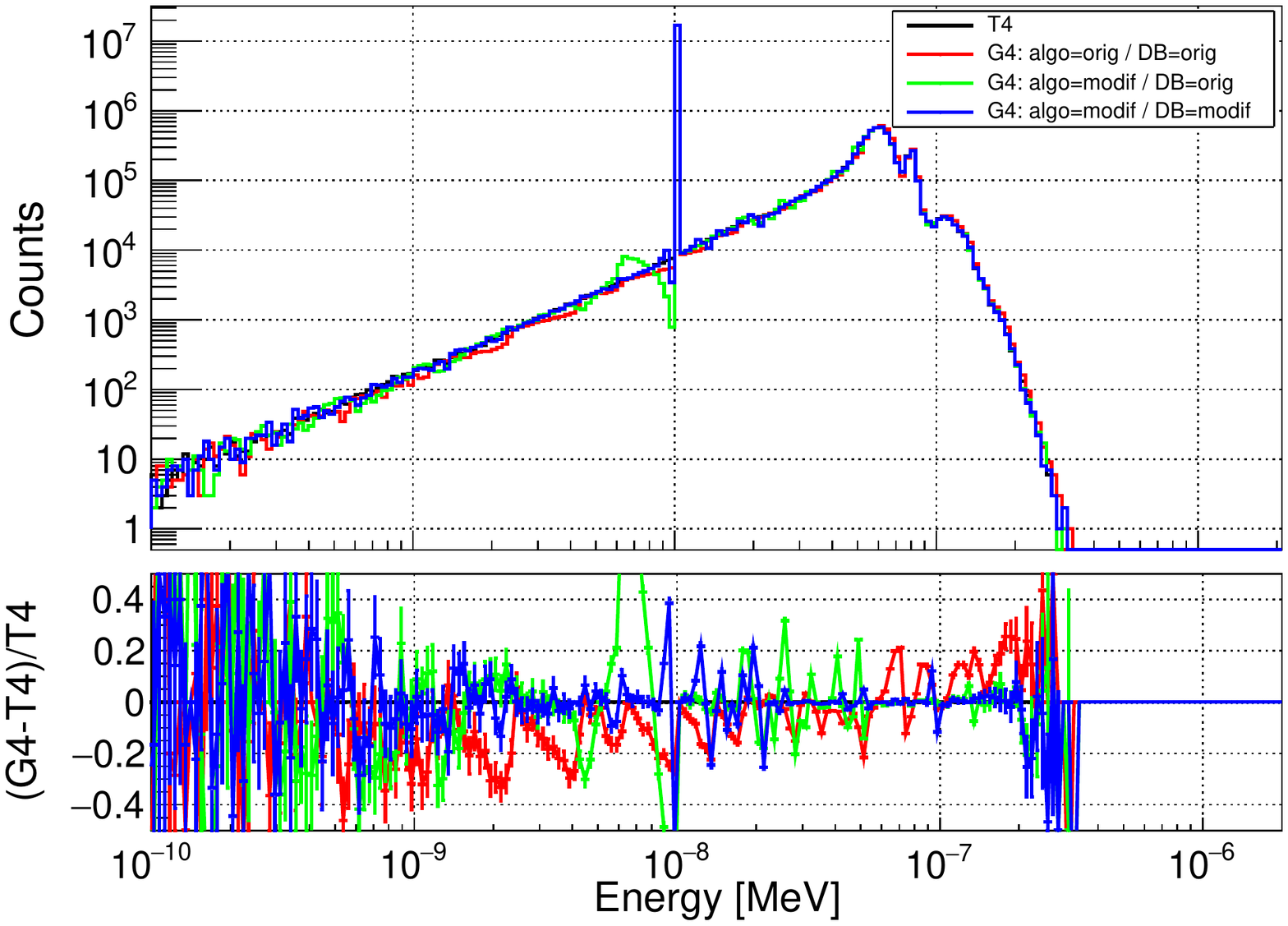}
         \caption{ENDF/B-VII.1 - Be metal TSL - 294K} 
     \end{subfigure}
     \hfill
     \begin{subfigure}[b]{0.49\linewidth}
         \centering
         \includegraphics[width=\linewidth]{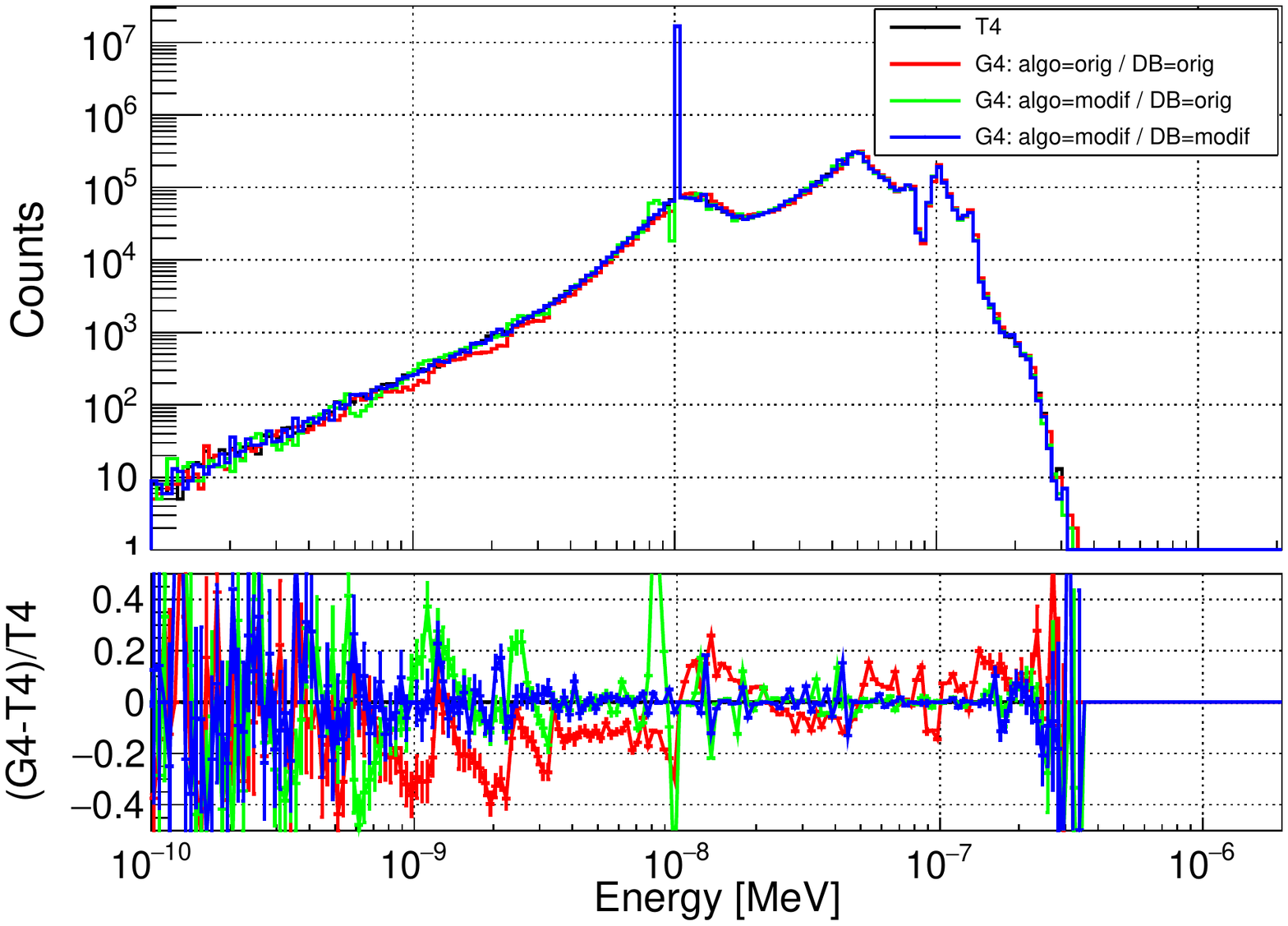}
         \caption{ENDF/B-VII.1 - BeO with BeinBeO and OinBeO TSL - 294K} 
     \end{subfigure}    

     \caption{\label{fig:B7_thinCylinder_energy_all} Scattered energy spectrum obtained with the thin cylinder benchmark with the ENDF/B-VII.1 library for different medium described by TSL data (top plot) and their relative differences using \tripoli~as the reference (bottom plot), for Geant4 original algorithm (red curve), Geant4 modified algorithm (green curve), Geant4 modified algorithm and reprocessed TSL data (blue curve)  and \tripoli~(black curve).}
\end{figure}

\begin{figure}[H]

    \centering
     \begin{subfigure}[b]{0.49\linewidth}
         \centering
         \includegraphics[width=\linewidth]{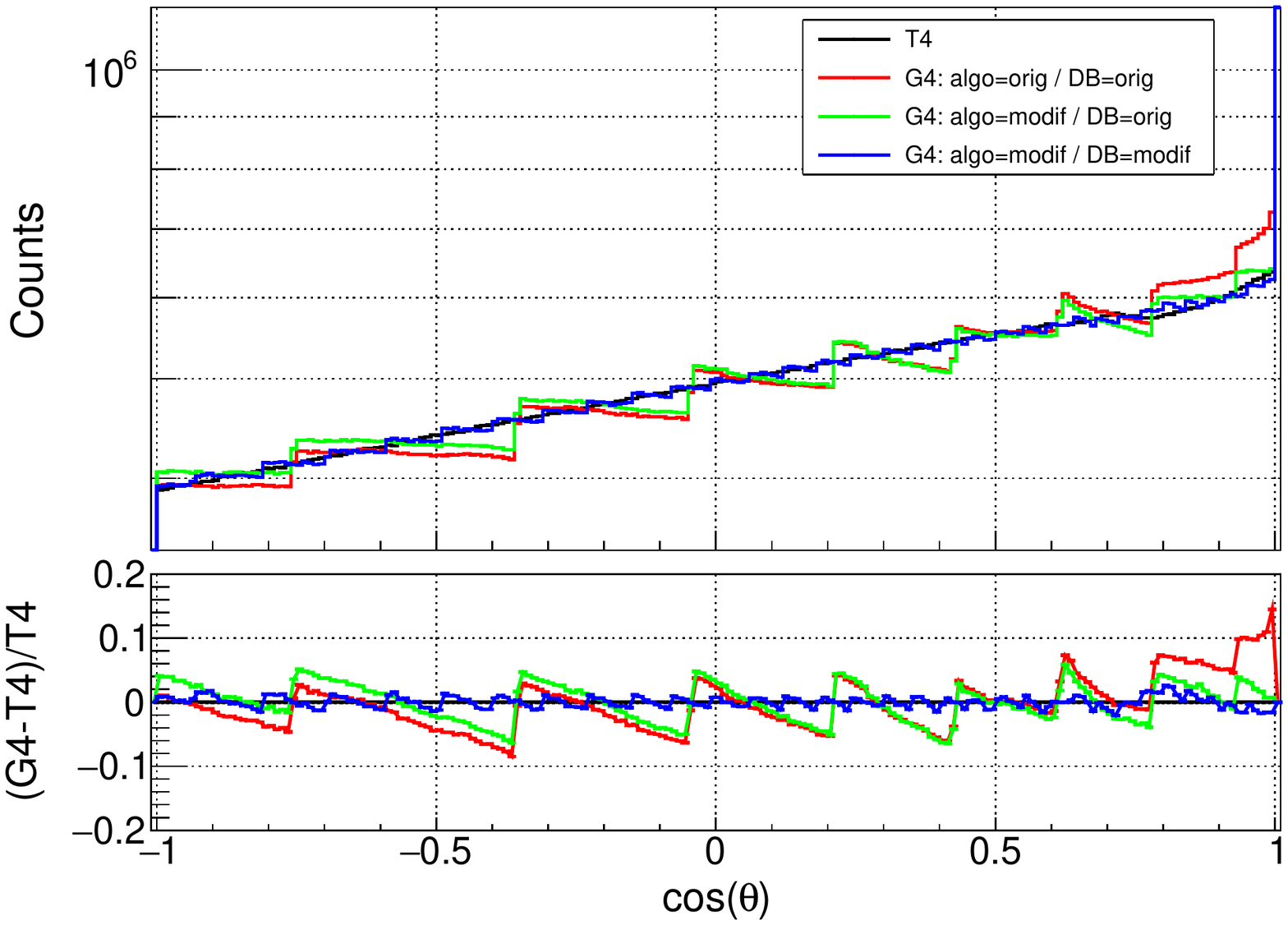}
         \caption{ENDF/B-VII.1 - CH$_2$ with HinCH$_2$ TSL - 296K}
     \end{subfigure}
     \hfill
     \begin{subfigure}[b]{0.49\linewidth}
         \centering
         \includegraphics[width=\linewidth]{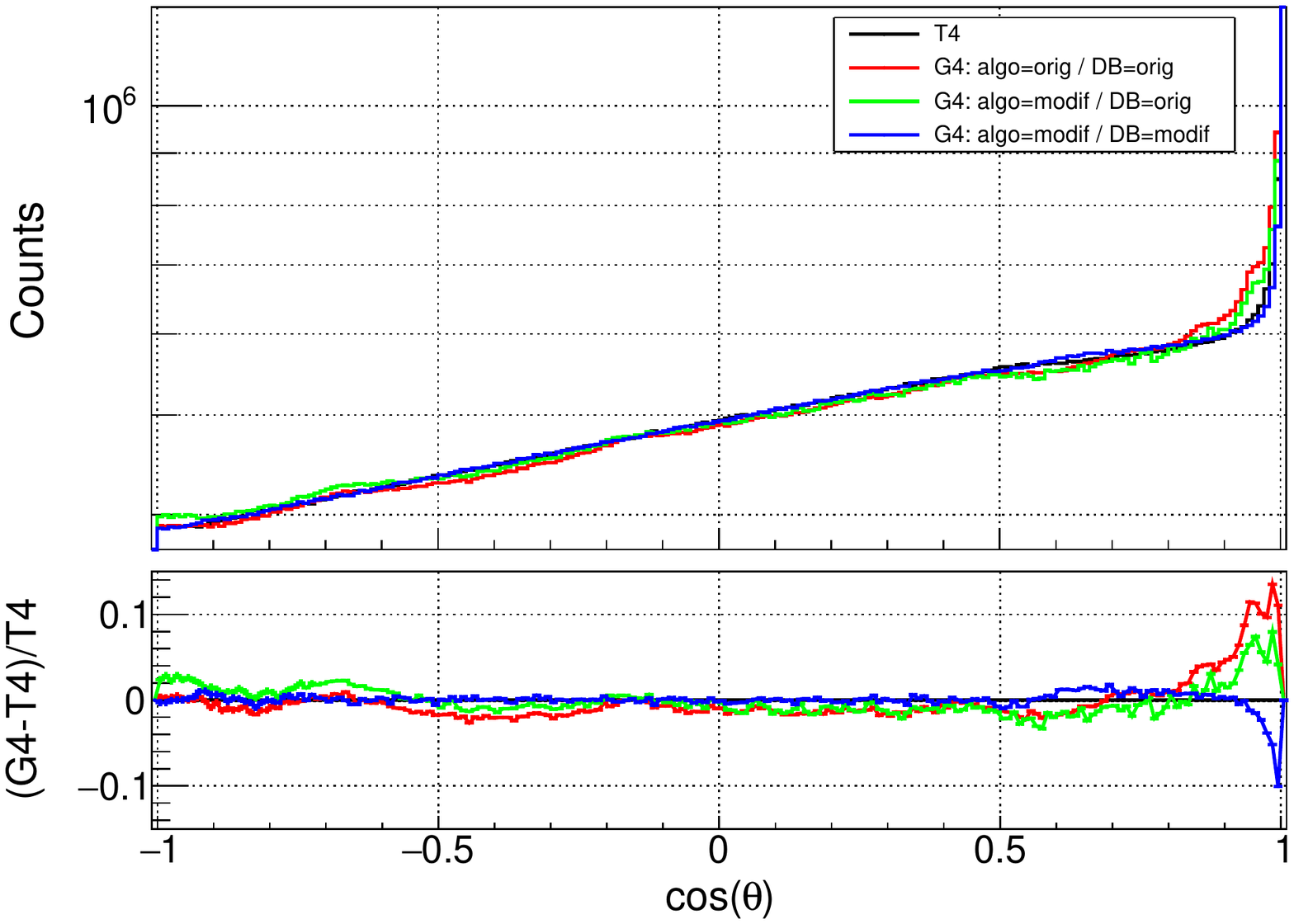}
         \caption{ENDF/B-VII.1 - H$_2$O with HinH$_2$O TSL - 294K}
     \end{subfigure}
     \hfill
     \begin{subfigure}[b]{0.49\linewidth}
         \centering
         \includegraphics[width=\linewidth]{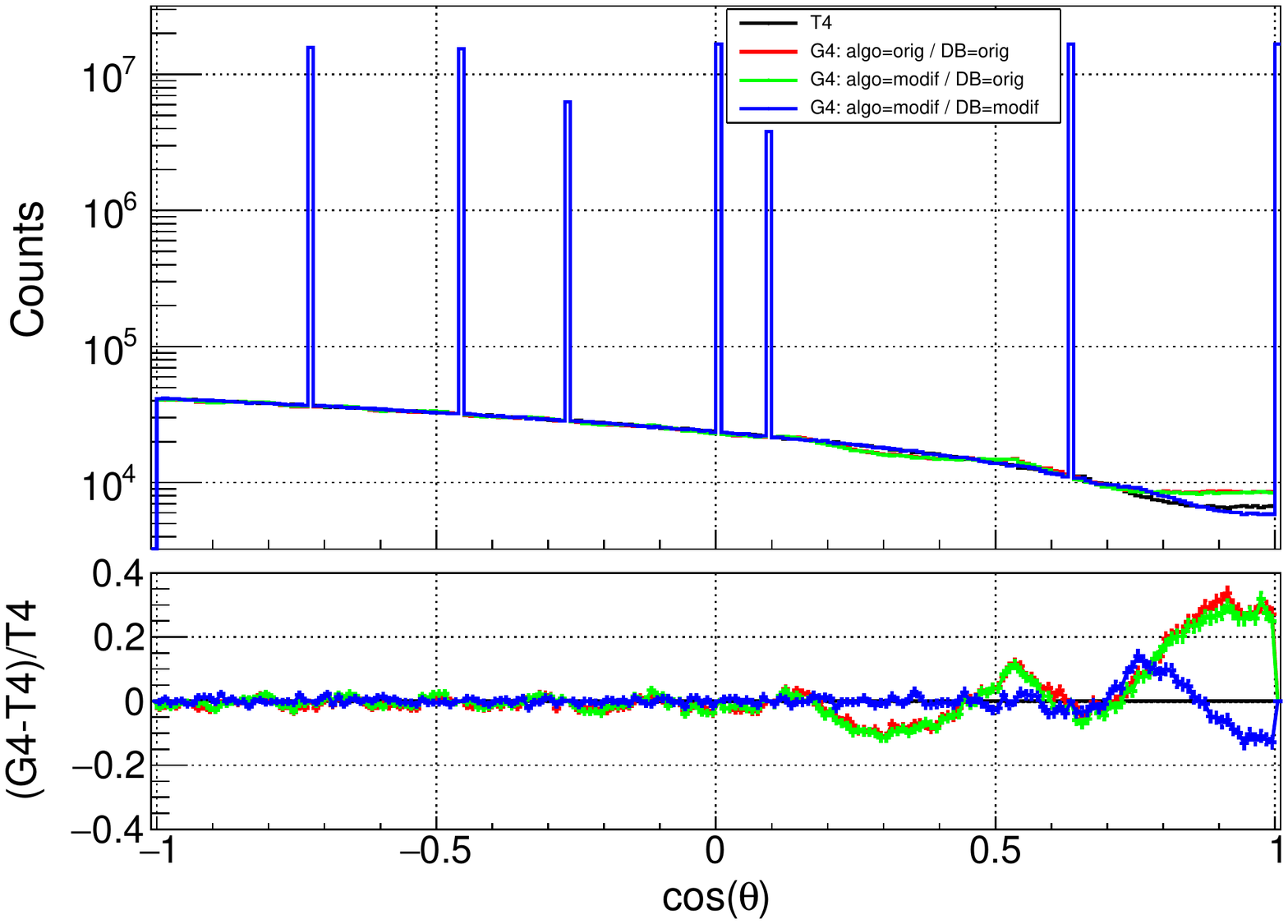}
         \caption{ENDF/B-VII.1 - Graphite TSL - 296K}
     \end{subfigure}
     \hfill
     \begin{subfigure}[b]{0.49\linewidth}
         \centering
         \includegraphics[width=\linewidth]{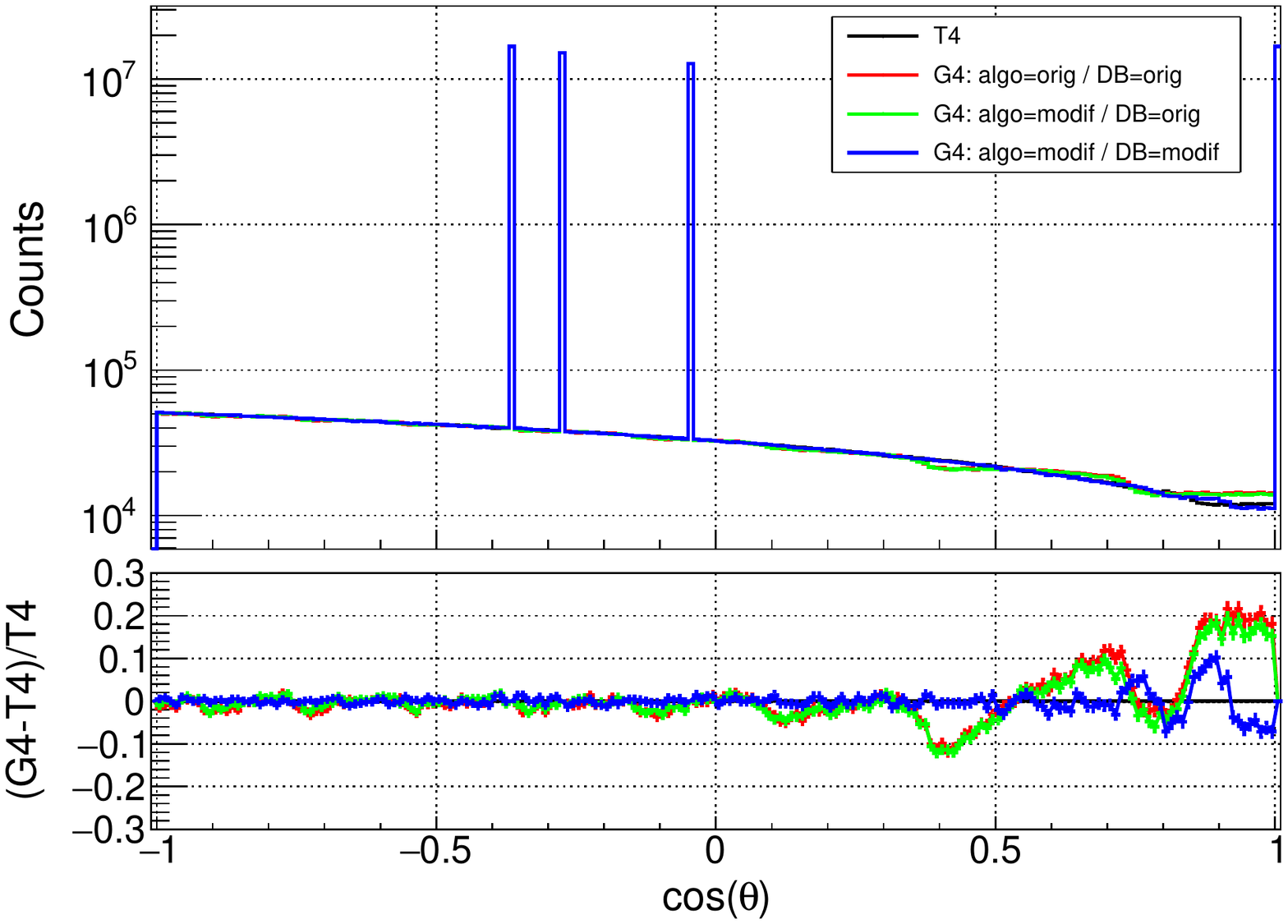}
         \caption{ENDF/B-VII.1 - Be metal TSL - 294K} 
     \end{subfigure}
     \hfill
     \begin{subfigure}[b]{0.49\linewidth}
         \centering
         \includegraphics[width=\linewidth]{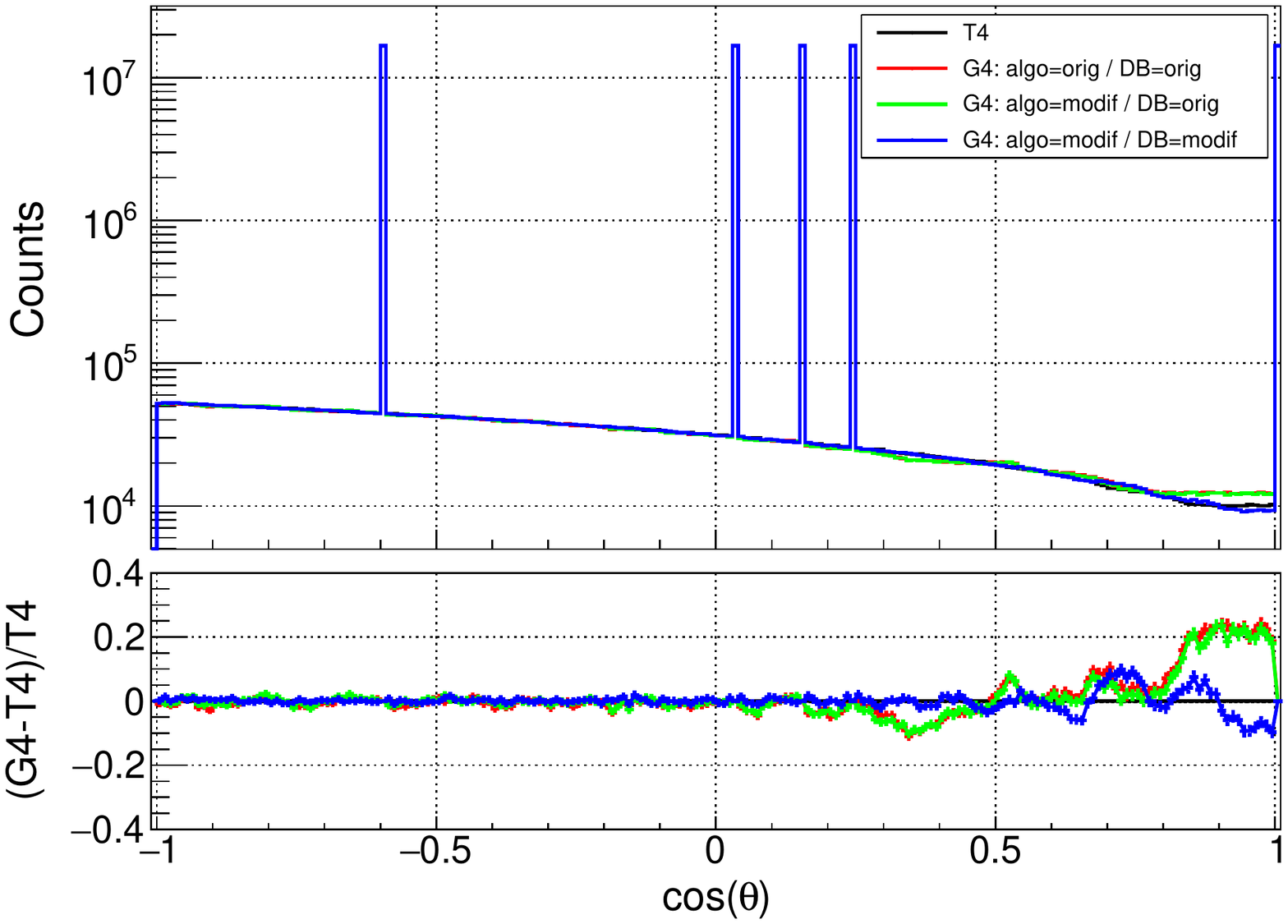}
         \caption{ENDF/B-VII.1 - BeO with BeinBeO and OinBeO TSL - 294K} 
     \end{subfigure}    
    
     \caption{\label{fig:B7_thinCylinder_cosTheta_all} Scattered energy cosinus angle obtained with the thin cylinder benchmark with the ENDF/B-VII.1 library for different medium described by TSL data (top plot) and their relative differences using \tripoli~as the reference (bottom plot), for Geant4 original algorithm (red curve), Geant4 modified algorithm (green curve), Geant4 modified algorithm and reprocessed TSL data (blue curve)  and \tripoli~(black curve).}
\end{figure}

In order to verify and validate this nuclear data processing tool, simulations were run using the "sphere" benchmark and the "thin-cylinder" benchmark with different moderator materials.
\newline
Figures \ref{fig:B7_sphere_all}, \ref{fig:B7_thinCylinder_energy_all} and \ref{fig:B7_thinCylinder_cosTheta_all} show that the Geant4 predictions obtained with ENDF/B-VII data processed by the new TSL data processing tool are in good agreement with old Geant4 predictions obtained with the TSL data distributed with Geant4 obtained with the following NJOY parameters: $tol$=0.02 and $N_{\mu}$=8 (DB=orig in figures). These values lead to the flux stair shapes visible between 10$^{-10}$ and 10$^{-8}$ MeV  in Figure \ref{fig:B7_thinCylinder_energy_all} and \ref{fig:B7_thinCylinder_cosTheta_all} especially for CH$_{2}$ (red and green curves). When modifying these parameters to $tol$=0.001 and $N_{\mu}$=32 (DB=modif in figures) this shape vanishes (blue curve). The remaining 'spikes' in the blue curves shown in Figure \ref{fig:B7_thinCylinder_cosTheta_all} would need further investigations.  

\clearpage

\clearpage

The advantage of having the ENDF/B-VIII.0 TSL data in Geant4 is that polyethylene as a cold moderator can be effectively studied since there is TSL data at 77 K while it was only available at 296 K in ENDF/B-VII.1.
Now this kind of study can be performed with the additional and updated materials in ENDF/B-VIII.0 and JEFF-3.3 (more focused on cold moderators) thermal libraries.

\begin{figure}[htbp]
    \begin{center}
	\includegraphics[scale=0.5]{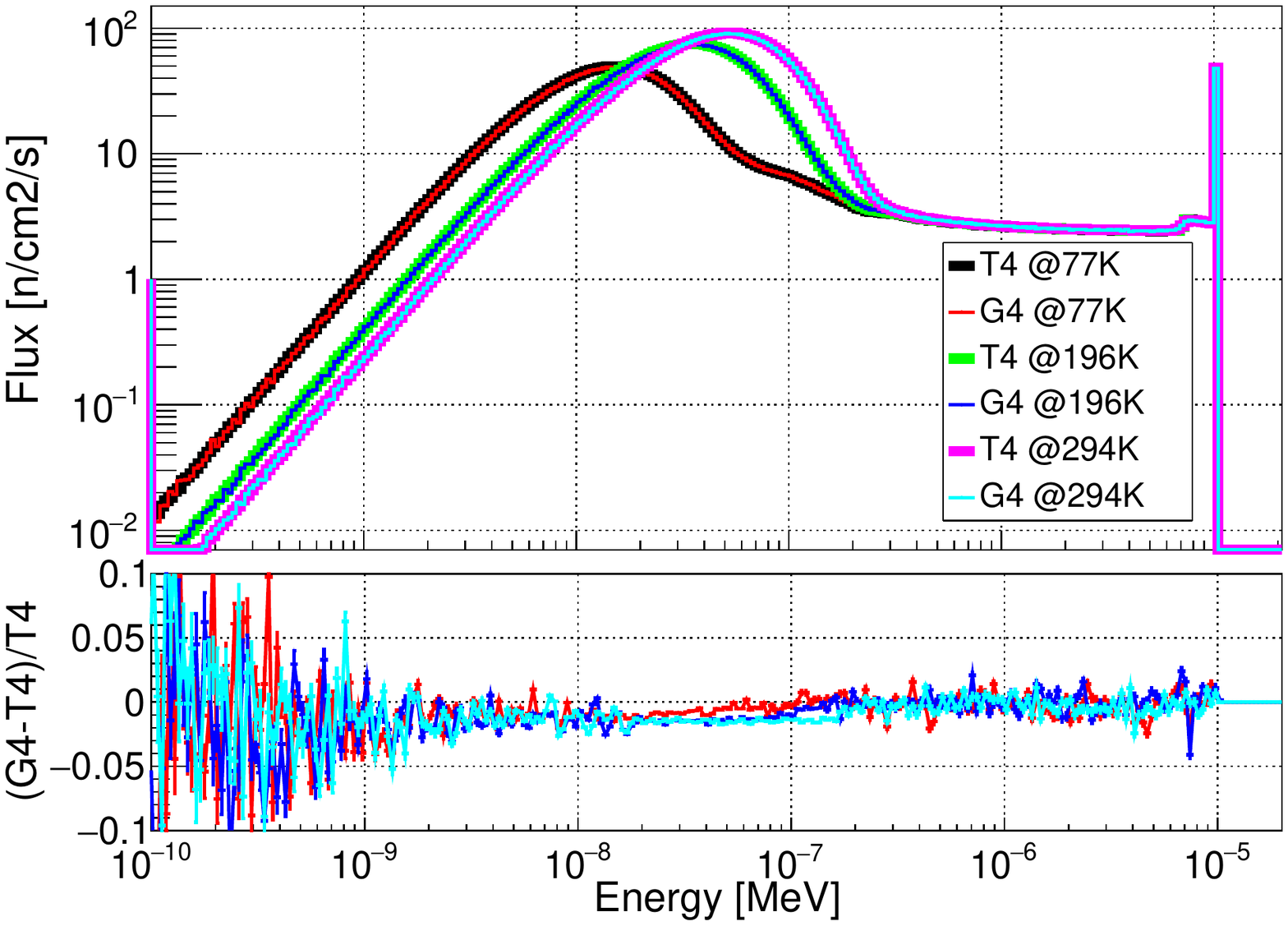}
	\end{center}
	\caption{\label{fig:B8_CH2_sphere} Neutron flux obtained with the sphere benchmark with the ENDF/B-VIII.0 library for a polyethylene (CH$_2$) medium as a function of the temperature (top plot) and their relative differences using \tripoli~as the reference (bottom plot).}
\end{figure}

\begin{figure}[htbp]
    \begin{center}
	\includegraphics[scale=0.5]{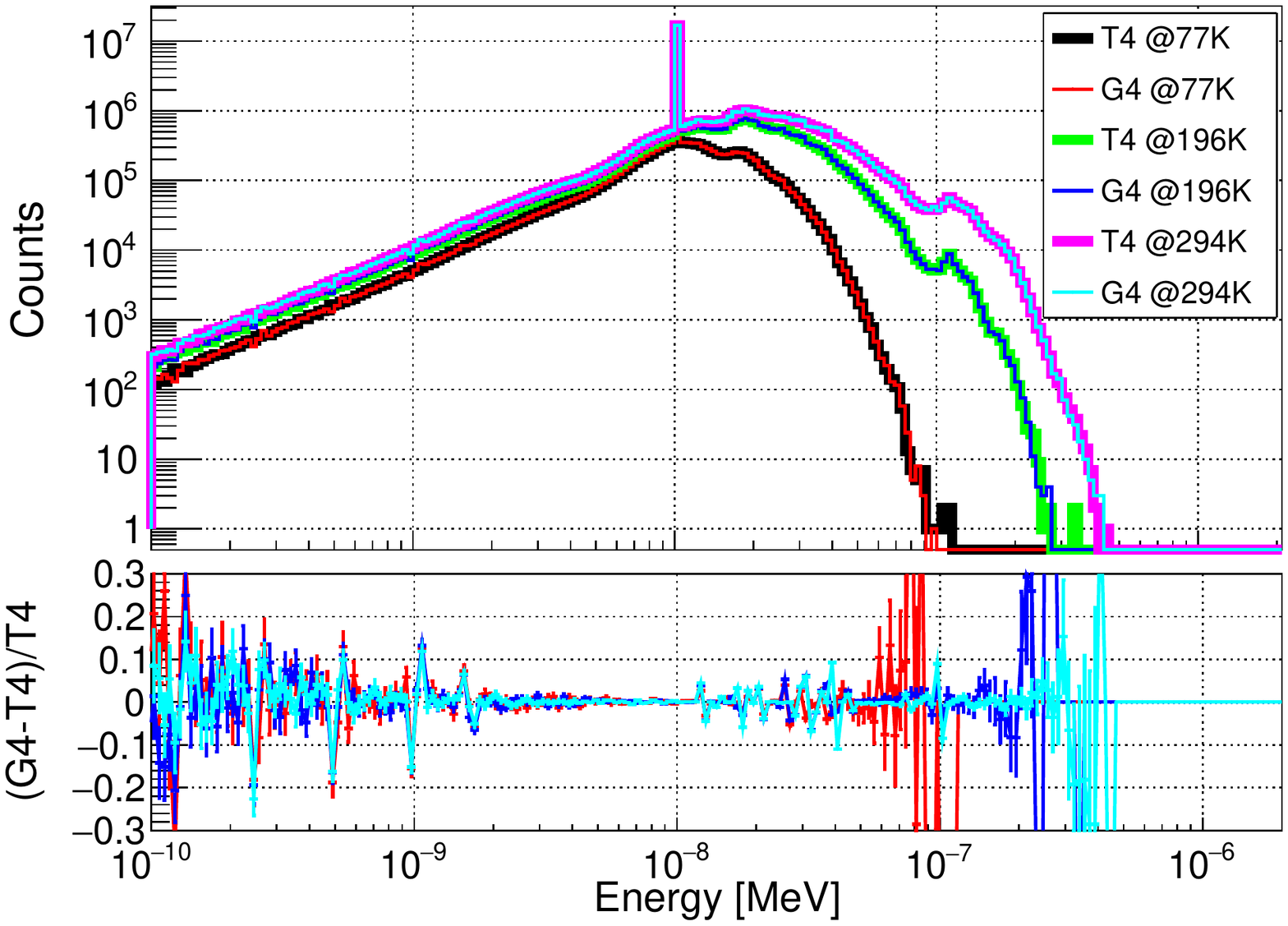}
	\end{center}
	\caption{\label{fig:B8_CH2_thinCylinder_energy} Scattered neutron energy spectrum obtained with the thin cylinder benchmark with the ENDF/B-VIII.0 library for a polyethylene (CH$_2$) medium as a function of the temperature (top plot) and their relative differences using \tripoli~as the reference (bottom plot).}
\end{figure}

\begin{figure}[htbp]
    \begin{center}
	\includegraphics[scale=0.5]{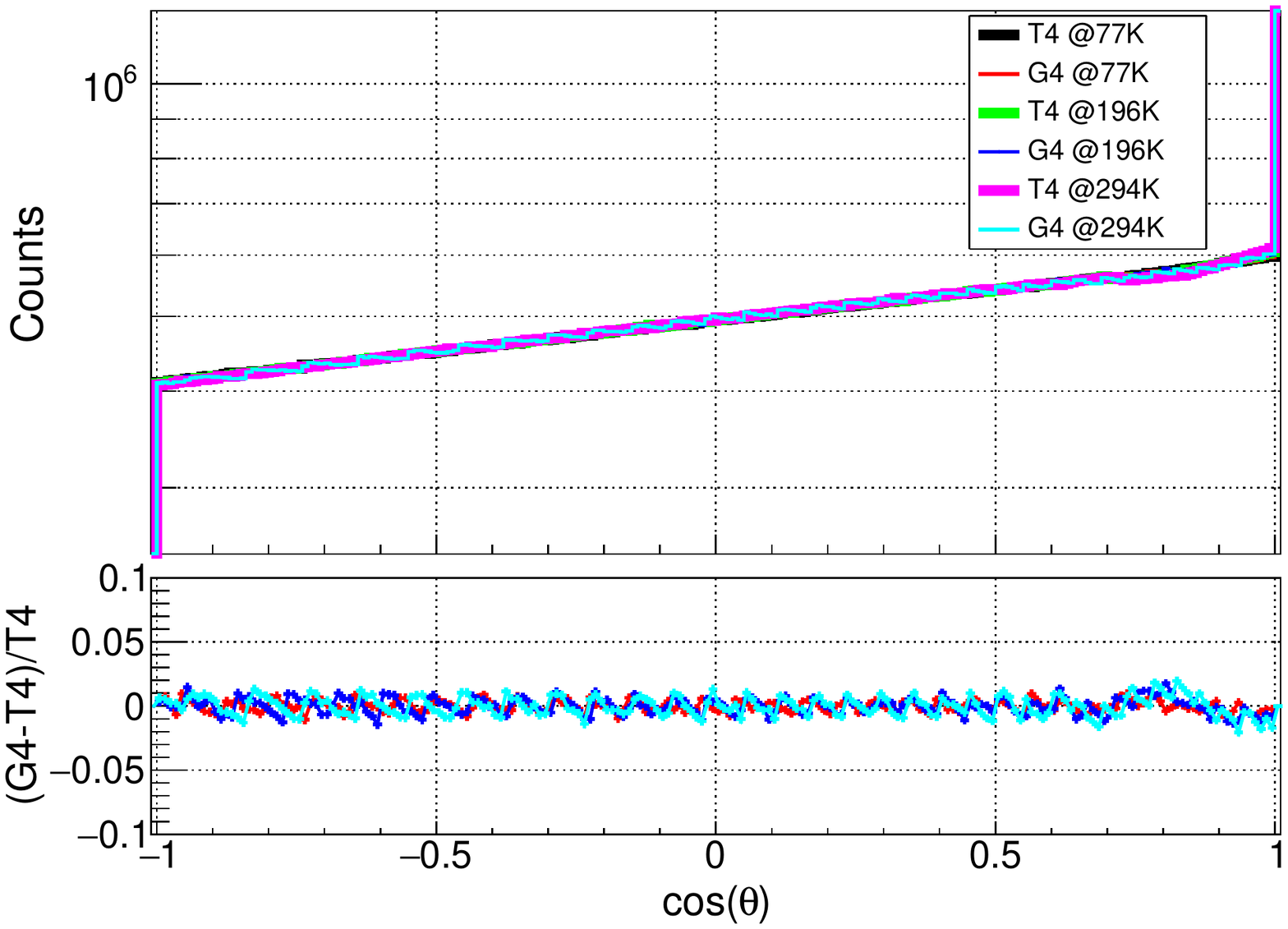}
	\end{center}
	\caption{\label{fig:B8_CH2_thinCylinder_cosTheta} Scattered neutron cosinus angle obtained with the thin cylinder benchmark with the ENDF/B-VIII.0 library for a polyethylene (CH$_2$) medium as a function of the temperature (top plot) and their relative differences using \tripoli~as the reference (bottom plot).}
\end{figure}

With this processing tool ENDF/B-VIII.0 TSL data have been integrated into Geant4 which allows having access to additional materials (e.g. YH$_{2}$, ice or SiC), to more material temperatures (e.g. at 77 K, 196 K, etc for CH$_2$ with ENDF-BVIII.0 instead of only 296 K with ENDF-BVII.1) and to newer TSL data compare to ENDF/B-VII.1 library. The same comparisons have been made between Geant4 and \tripoli~for ENDF/B-VIII.0 and are presented in Figures \ref{fig:B8_CH2_sphere}, \ref{fig:B8_CH2_thinCylinder_energy} and \ref{fig:B8_CH2_thinCylinder_cosTheta}.  Here the focus is made on polyethylene (CH$_2$) at different temperatures since it is often used as a thermal and cold moderator in numerous applications because it is easy to handle. 

\subsection{Geant4 accuracy and speed improvements}

In this work three improvements where made to the Geant4 code (labelled "algo=modif" in figures) compared to Geant4 version 10.07.p01 (algo=orig in figures), in particular in \textit{G4ParticleHPThermalScattering} and \textit{G4ParticleHPElastic} classes, to reduce Geant4/\tripoli~discrepancies greater than 20 \%, which can be observed on the red curves of Figure \ref{fig:B7_sphere_all}. For example in Figure \ref{fig:B7_sphere_all} for polyethylene or light water, (1) a flux discontinuity appears at 4 eV between Geant4 and \tripoli~and (2) below 10$^{-7}$ MeV there is firstly an overestimation followed by an underestimation of the Geant4 flux compared to \tripoli~greater than 20 \%, this behaviour is named the "wave shape" in the following. 
\newline
\indent
The cross-section discontinuity visible at 4 eV occurs at the energy where the transition between the nuclear cross-section (above 4 eV) and thermal scattering data (below 4 eV) occurs. The continuity between the two domains is ensured by NJOY and there is no problem in \tripoli. This shows that TSL data in Geant4 is not properly handled in the \textit{G4ParticleHPThermalScattering} class. In fact when a compound material (e.g. CH$_{2}$) is made of one element described by TSL data (e.g. H in CH$_{2}$) and the other by the free gas approximation (e.g. C in CH$_{2}$), if the neutron interacts with the nucleus described by the free gas approximation (e.g. C in CH$_{2}$), the \textit{G4ParticleHPElastic} class is called. However in this class, the target nucleus (e.g. H or C) is sampled again even if the target is already known (e.g. C). This modifies the interaction probability in giving to the nucleus described by TSL data more chance to be the target (e.g. H is more sampled than C in CH$_{2}$). After solving this problem, the Geant4 predictions represented by the green and blue curves are obtained in Figure \ref{fig:B7_sphere_all} and show that an accuracy below 2 \% is achieved around the 4 eV transition.
\newline
\indent
The "wave shape" visible between 10$^{-9}$ and 10$^{-7}$ MeV in Figure \ref{fig:B7_sphere_all} (red curve) contributing to large discrepancies (20\%) comes mainly from interpolation inaccuracies. For inelastic collisions, the outgoing energy probability $P(E\rightarrow E'_{i})$ is tabulated and computed with the cross-section $\sigma (E\rightarrow E'_{i})$ and the final energy bin width $\Delta E'_{i}=E'_{i,high}-E'_{i,low}$ such as: 

\begin{equation}
  P(E\rightarrow E'_{i})=\sigma(E\rightarrow E'_{i,low})\Delta E'_{i} 
\end{equation}\label{eq:SEsampling_old}
\noindent
However in case of large linear increasing cross-section the probability can be underestimated. Therefore the averaged cross-section over the bin width is used instead which is given by:

\begin{equation}
    P(E\rightarrow E'_{i})=\frac{\sigma(E\rightarrow E'_{i,low}) + \sigma(E\rightarrow E'_{i,up})}{2}\Delta E'_{i}
\end{equation}\label{eq:SEsampling_new}

\noindent
Once an outgoing energy bin has been sampled, a linear interpolation is made between the available secondary energy inside the bin. With these modifications removing this "wave shape", Geant4 agrees well with \tripoli~to better than 2 \% (green and blue curves). The remaining small discrepancy could be solved by investigating other interpolation methods used in Geant4.
\newline 
\indent
Concerning computing times, reference codes such as MCNP or \tripoli~seem to offer sensitively increased figures of merit when compared to Geant4. In closely looking at Geant4 TSL treatment, a lot of unnecessary interpolations are made, the most striking example being the temperature interpolation that is made when computing the neutron outgoing characteristics. In fact to get the neutron characteristics, Geant4 looks at the material temperature $T_{mat}$ and finds the corresponding temperature bin defined by its limits $T_{i}$ and $T_{i+1}$ for which a final state is sampled. Then the neutron final state at $T_{mat}$ is computed by a linear temperature interpolation. This is done even if $T_{mat}=T_{i}$. The main problem is that the temperature interpolation should be performed stochastically \cite{Donnelly2011InterpolationOT}, \textit{i.e.} the temperature $T_{i}$ or  $T_{i+1}$ should be randomly selected and then the final state should be computed at the sampled temperature. Physically this is the right method to use because the quantity that can only be interpolated is the phonon spectrum from which NJOY makes the convolution process to get TSL data, not the outgoing neutron characteristics.  
In addition to dealing with the TSL data in the right way, using a stochastic temperature interpolation speeds up the code by a factor of two since now only one final state is sampled instead of two at each simulation step.
\newline 
\indent
Overall these modifications allow reducing the Geant4/\tripoli~discrepancies from 20\% to less than 2\% as seen in Figure \ref{fig:B7_sphere_all} and increase the speed by at least a factor of two.

\section{Conclusion}
This work presented improvements that should be taken into account in the next Geant4 release to make Neutron-HP package on-par with reference neutron transport codes such as \tripoli~or MCNP based on accuracy and speed criteria. Geant4 limitations have been overcome in implementing the SVT algorithm allowing the conservation of the average thermal reaction rate when the free gas approximation is used and improving and debugging the methods used to sample the thermal neutron outgoing characteristics after one collision from TSL data. These reduce Geant4/\tripoli~discrepancies from as much as 20\% to less than 2\% and increase Geant4 code speed by at least a factor of two. The remaining 2 \% discrepancy could come from numerous numerical aspects (interpolation methods, etc.) which will require further investigations. In the future the DBRC treatment \cite{Becker2009} of the neutron elastic scattering and the use of probability table to describe cross-sections in the unresolved resonance region should be taken into account in Geant4 to be fully competitive with "state-of-the-art" neutron transport codes.
\newline
A TSL nuclear data processing tool has also been developed and validated to take into account in Geant4 new evaluated nuclear data libraries such as ENDF/B-VIII.0 and JEFF-3.3 instead of being limited only to ENDF/B-VII.1 evaluation dating back to 2011. This allows the users to test a broader range of materials when designing experiments.

\subsection*{Acknowledgements}
The authors wish to sincerely thank V. Jaiswal from IRSN for his substantial help in understanding the S($\alpha,\beta$) format of ENDF. \tripoli~is a registered trademark of CEA. The authors thank Electricité de France (EDF) for partial financial support.


\begin{thebibliography}{00}
\providecommand{\natexlab}[1]{#1}
\providecommand{\url}[1]{\texttt{#1}}
\expandafter\ifx\csname urlstyle\endcsname\relax
  \providecommand{\doi}[1]{doi: #1}\else
  \providecommand{\doi}{doi: \begingroup \urlstyle{rm}\Url}\fi

\bibitem[Lux and Koblinger(1991)]{Lux:268101}
Iván Lux and László Koblinger.
\newblock \emph{{Monte Carlo particle transport methods: neutron and photon
  calculations}}.
\newblock CRC Press, Boca Raton, FL, 1991.
\newblock URL \url{https://cds.cern.ch/record/268101}.

\bibitem[Becker et~al.(2009{\natexlab{a}})Becker, Dagan, and
  Lohnert]{BECKER2009470}
B.~Becker, R.~Dagan, and G.~Lohnert.
\newblock Proof and implementation of the stochastic formula for ideal gas,
  energy dependent scattering kernel.
\newblock \emph{Annals of Nuclear Energy}, 36\penalty0 (4):\penalty0 470--474,
  2009{\natexlab{a}}.
\newblock ISSN 0306-4549.
\newblock \doi{https://doi.org/10.1016/j.anucene.2008.12.001}.
\newblock URL
  \url{https://www.sciencedirect.com/science/article/pii/S0306454908003186}.

\bibitem[et~al.(2011)]{Chadwick2011}
M.~B.~Chadwick et~al.
\newblock {ENDF/B-VII.1 nuclear data for science and technology: Cross
  sections, covariances, fission product yields and decay data}.
\newblock \emph{Nuclear Data Sheets}, 112:\penalty0 2887--2996, 12 2011.
\newblock ISSN 00903752.
\newblock \doi{10.1016/j.nds.2011.11.002}.

\bibitem[et~al.(2018{\natexlab{a}})]{Brown2018}
D.A.~Brown et~al.
\newblock {ENDF/B-VIII.0: The 8th Major Release of the Nuclear Reaction Data
  Library with CIELO-project Cross Sections, New Standards and Thermal
  Scattering Data}.
\newblock \emph{Nuclear Data Sheets}, 148:\penalty0 1--142, feb
  2018{\natexlab{a}}.
\newblock ISSN 0090-3752.
\newblock \doi{10.1016/J.NDS.2018.02.001}.
\newblock URL
  \url{https://www.sciencedirect.com/science/article/pii/S0090375218300206}.

\bibitem[operation Volume~42(2020)]{JEFF-TSL2020}
International Evaluation~Co operation Volume~42.
\newblock {Thermal Scattering Law S($\alpha$,$\beta$): Measurement, Evaluation
  and Application}.
\newblock Technical report, NUCLEAR ENERGY AGENCY ORGANISATION FOR ECONOMIC
  CO-OPERATION AND DEVELOPMENT, 2020.

\bibitem[Team(2003)]{mcnp5}
X-5 Monte~Carlo Team.
\newblock {MCNP - Version 5, Vol. I: Overview and Theory}.
\newblock \emph{LA-UR-03-1987}, 2003.

\bibitem[et~al.(2012)]{mcnp6}
T.~Goorley et~al.
\newblock {Initial MCNP6 Release Overview}.
\newblock \emph{Nuclear Technology}, 180:\penalty0 298--315, 2012.

\bibitem[et~al.(2015{\natexlab{a}})]{scale}
B.T.~Rearden et~al.
\newblock {Monte Carlo capabilities of the SCALE code system}.
\newblock \emph{Annals of Nuclear Energy}, 82:\penalty0 130--141,
  2015{\natexlab{a}}.

\bibitem[et~al.(2015{\natexlab{b}})]{serpent}
J.~Leppänen et~al.
\newblock {The Serpent Monte Carlo code: Status, development and applications
  in 2013}.
\newblock \emph{Annals of Nuclear Energy}, 82:\penalty0 298--315,
  2015{\natexlab{b}}.

\bibitem[et~al.(2015{\natexlab{c}})]{moret5}
O.~Jacquet et~al.
\newblock {Capabilities overview of the MORET 5 Monte Carlo code}.
\newblock \emph{Annals of Nuclear Energy}, 82:\penalty0 74--84,
  2015{\natexlab{c}}.

\bibitem[et~al.(2021)]{moret6}
Eric~Dumonteil et~al.
\newblock {Patchy nuclear chain reactions}.
\newblock \emph{Communications Physics 2021 4:1}, 4:\penalty0 1--10, 7 2021.
\newblock ISSN 2399-3650.
\newblock \doi{10.1038/s42005-021-00654-9}.
\newblock URL \url{https://www.nature.com/articles/s42005-021-00654-9}.

\bibitem[et~al(2015)]{tripoli4}
E.~Brun et~al.
\newblock {TRIPOLI-4, CEA, EDF and AREVA reference Monte Carlo code}.
\newblock \emph{Annals of Nuclear Energy}, 82:\penalty0 151–160, 2015.

\bibitem[et~al.(2019)]{intercomp}
I.~Duhamel et~al.
\newblock {International Criticality Benchmark Comparison for Nuclear Data
  Validation}.
\newblock \emph{Transactions of the American Nuclear Society}, 121, 2019.

\bibitem[et~al.(1978)]{geant}
R.~Brun et~al.
\newblock {Simulation program for particle physics experiments, GEANT: user
  guide and reference manual}.
\newblock \emph{CERN Report CERN-DD-78-2}, 1978.

\bibitem[et~al.(2016)]{Allison2016}
J.~Allison et~al.
\newblock {Recent developments in Geant4}.
\newblock \emph{Nuclear Instruments and Methods in Physics Research Section A:
  Accelerators, Spectrometers, Detectors and Associated Equipment},
  835:\penalty0 186--225, nov 2016.
\newblock ISSN 0168-9002.
\newblock \doi{10.1016/J.NIMA.2016.06.125}.
\newblock URL
  \url{https://www.sciencedirect.com/science/article/pii/S0168900216306957}.

\bibitem[et~al.(2014{\natexlab{a}})]{Mendoza2014}
E.~Mendoza et~al.
\newblock {New standard evaluated neutron cross section libraries for the
  GEANT4 code and first verification}.
\newblock \emph{IEEE Transactions on Nuclear Science}, 61:\penalty0 2357--2364,
  2014{\natexlab{a}}.
\newblock ISSN 00189499.
\newblock \doi{10.1109/TNS.2014.2335538}.

\bibitem[et~al.(2018{\natexlab{b}})]{Hartling2018}
K.~Hartling et~al.
\newblock {The effects of nuclear data library processing on Geant4 and MCNP
  simulations of the thermal neutron scattering law}.
\newblock \emph{Nuclear Instruments and Methods in Physics Research, Section A:
  Accelerators, Spectrometers, Detectors and Associated Equipment},
  891:\penalty0 25--31, 5 2018{\natexlab{b}}.
\newblock ISSN 01689002.
\newblock \doi{10.1016/j.nima.2018.02.053}.

\bibitem[et~al.(2018{\natexlab{c}})]{Tran2018a}
H.N.~Tran et~al.
\newblock {Comparison of the thermal neutron scattering treatment in MCNP6 and
  GEANT4 codes}.
\newblock \emph{Nuclear Instruments and Methods in Physics Research Section A:
  Accelerators, Spectrometers, Detectors and Associated Equipment},
  893:\penalty0 84--94, jun 2018{\natexlab{c}}.
\newblock ISSN 01689002.
\newblock \doi{10.1016/j.nima.2018.02.094}.
\newblock URL
  \url{https://linkinghub.elsevier.com/retrieve/pii/S0168900218302651}.

\bibitem[Mendoza et~al.(2018)Mendoza, Cano-Ott, and Capote]{Mendoza2018}
E.~Mendoza, D.~Cano-Ott, and R.~Capote.
\newblock { Update of the Evaluated Neutron Cross Section Libraries for the
  Geant4 Code, IAEA technical report INDC(NDS)-0758 (releases JEFF-3.3,
  JEFF-3.2, ENDF/B-VIII.0, ENDF/B-VII.1, BROND-3.1 and JENDL-4.0u)}.
\newblock Technical report, CIEMAT, Vienna, 2018.
\newblock URL
  \url{{https://www-nds.iaea.org/geant4/figures/G4\_10.04.p01\_VS\_MCNP6\_ENDF80.pdf}}.

\bibitem[Coveyou et~al.(1956)Coveyou, Bate, and Osborn]{Coveyou1956}
R.~R. Coveyou, R.~R. Bate, and R.~K. Osborn.
\newblock Effect of moderator temperature upon neutron flux in infinite,
  capturing medium.
\newblock \emph{Journal of Nuclear Energy}, 2:\penalty0 153--167, 1 1956.
\newblock ISSN 08913919.
\newblock \doi{10.1016/0891-3919(55)90030-9}.

\bibitem[Agency(2020{\natexlab{a}})]{icsbep}
OECD / Nuclear~Energy Agency.
\newblock {International Criticality Safety Benchmark Evaluation Project
  (ICSBEP)}.
\newblock
  \url{https://www.oecd-nea.org/jcms/pl\_24498/international-criticality-safety-benchmark-evaluation-project-icsbep},
  2020{\natexlab{a}}.

\bibitem[et~al.(2014{\natexlab{b}})]{blaise}
P.~Blaise et~al.
\newblock {Monte Carlo Modelling of Increasing Void Fraction in 100\% MOX ABWR:
  Lessons Drawn from the FUBILA Program}.
\newblock \emph{J. Nucl. Sci. Technol.}, 47:\penalty0 558--569,
  2014{\natexlab{b}}.

\bibitem[et~al.(2010)]{fausser}
C.~Fausser et~al.
\newblock {Numerical Benchmarks TRIPOLI-MCNP with use of MCAM on FNG ITER Bulk
  Shield and FNG HCLL TBM mock-up experiments.}
\newblock \emph{Proceedings of the SOFT 2010 Conference, Porto, Portugal},
  2010.

\bibitem[de~Sûreté~Nucléaire(2017)]{asnguide28}
Autorité de~Sûreté~Nucléaire.
\newblock {Guide de l'ASN n°28 : Qualification des outils de calcul
  scientifique utilisés dans la démonstration de sûreté nucléaire}.
\newblock
  \url{https://www.asn.fr/content/download/118129/1028436/version/2/file/Guide\%2028\%20QOCS.pdf},
  2017.

\bibitem[Everett and Cashwell(1983)]{Everett1983}
C.~J. Everett and E.~D. Cashwell.
\newblock {A Third Monte Carlo Sampler (A Revision and Extension of Samplers I
  and 11), LA-9721-MS, UC-32}.
\newblock Technical report, Los Alamos National Laboratory (LANL), 1983.
\newblock URL
  \url{https://laws.lanl.gov/vhosts/mcnp.lanl.gov/pdf\_files/la-9721.pdf}.

\bibitem[Becker et~al.(2009{\natexlab{b}})Becker, Dagan, and
  Lohnert]{Becker2009}
B.~Becker, R.~Dagan, and G.~Lohnert.
\newblock {Proof and implementation of the stochastic formula for ideal gas,
  energy dependent scattering kernel}.
\newblock \emph{Annals of Nuclear Energy}, 36\penalty0 (4):\penalty0 470--474,
  may 2009{\natexlab{b}}.
\newblock ISSN 03064549.
\newblock \doi{10.1016/j.anucene.2008.12.001}.

\bibitem[et~al.(2014{\natexlab{c}})]{zoiadbrc1}
A.~Zoia et~al.
\newblock {Doppler broadening of neutron elastic scattering kernel in
  TRIPOLI-4}.
\newblock \emph{Annals of Nuclear Energy}, 54:\penalty0 218--226,
  2014{\natexlab{c}}.
\newblock \doi{https://10.1016/j.anucene.2012.11.023}.

\bibitem[Squires(2012)]{Squires2012}
G.~L. Squires.
\newblock {Introduction to the theory of Thermal Neutron Scattering}.
\newblock \emph{Cambridge University Press}, 2012.

\bibitem[et~al.(2005)]{gromacs}
D.~Van Der~Spoel et~al.
\newblock {GROMACS: fast, flexible, and free.}
\newblock \emph{Journal of Computational Chemistry}, 47:\penalty0 1701–1718,
  2005.
\newblock \doi{https://doi.org/10.1002/jcc.20291}.

\bibitem[Kresse and Furthmüller(1996)]{Kresse1996}
G.~Kresse and J.~Furthmüller.
\newblock Efficient iterative schemes for ab initio total-energy calculations
  using a plane-wave basis set.
\newblock \emph{Physical Review B}, 54:\penalty0 11169, 10 1996.
\newblock \doi{10.1103/PhysRevB.54.11169}.
\newblock URL
  \url{https://journals.aps.org/prb/abstract/10.1103/PhysRevB.54.11169}.

\bibitem[Agency(2020{\natexlab{b}})]{neatsl}
OECD / Nuclear~Energy Agency.
\newblock {SG42}.
\newblock
  \url{https://www.oecd-nea.org/upload/docs/application/pdf/2020-03/volume42.pdf},
  2020{\natexlab{b}}.

\bibitem[et~al.(2017)]{Macfarlane2017}
R.~Macfarlane et~al.
\newblock {The NJOY Nuclear Data Processing System, Version 2016}.
\newblock Technical report, Los Alamos National Laboratory (LANL), Los Alamos,
  NM (United States), jan 2017.
\newblock URL \url{http://www.osti.gov/servlets/purl/1338791/}.

\bibitem[Donnelly(2011)]{Donnelly2011InterpolationOT}
J.~Donnelly.
\newblock {Interpolation of temperature-dependent nuclide data in MCNP}.
\newblock \emph{Nuclear Science and Engineering}, 168:\penalty0 180 -- 184,
  2011.
  
  \bibitem[Thulliez(2021)]{GitLab}
  L.~Thulliez et~al.
  \newblock {geant4\_tsl\_processing code}.
\newblock URL \url{https://gitlab.com/lthullie/geant4\_tsl\_processing}

\end{thebibliography}
\end{document}